\documentclass[12pt]{article}

\usepackage{amsmath,amsfonts}
\usepackage{bm}
\usepackage{natbib,theorem}
\usepackage{graphicx}
\usepackage{amssymb}
\usepackage{latexsym}
\usepackage{multicol}
\usepackage{color}
\usepackage{setspace}
\usepackage{hyperref}

\renewcommand{\baselinestretch} {1.30}
\newtheorem{theorem}{\bf Theorem}
\newtheorem{lemma}{\bf Lemma}
\newtheorem{corollary}{\bf Corollary}

\theoremheaderfont{\hskip\parindent\normalfont}
\theorembodyfont{\normalfont}

\newcommand{\bA}{\mathbf{A}}
\newcommand{\ba}{\mathbf{a}}
\newcommand{\bB}{\mathbf{B}}

\newcommand{\bD}{\mathbf{D}}

\newcommand{\bI}{\mathbf{I}}

\newcommand{\bU}{\mathbf{U}}
\newcommand{\bV}{\mathbf{V}}

\newcommand{\bW}{\mathbf{W}}
\newcommand{\bw}{\mathbf{w}}
\newcommand{\bX}{\mathbf{X}}
\newcommand{\bx}{\mathbf{x}}
\newcommand{\bY}{\mathbf{Y}}

\newcommand{\cN}{\mathcal{N}}
\newcommand{\cX}{\mathcal{X}}

\newcommand{\bbeta}{\boldsymbol{\beta}}
\newcommand{\bdelta}{\boldsymbol{\delta}}
\newcommand{\bDelta}{\boldsymbol{\Delta}}

\newcommand{\bomega}{\boldsymbol{\omega}}
\newcommand{\bmu}{\boldsymbol{\mu}}
\newcommand{\bnu}{\boldsymbol{\nu}}
\newcommand{\bSigma}{\boldsymbol{\Sigma}}

\newcommand{\bxi}{\boldsymbol{\xi}}

\newcommand{\bzero}{\boldsymbol{0}}
\newcommand{\bone}{\boldsymbol{1}}

\begin{document}

\numberwithin{equation}{section}
\renewcommand{\baselinestretch}{1.5}

\title{\bf Sparsifying the Fisher Linear Discriminant by Rotation}

\author{Ning Hao, Bin Dong, and Jianqing Fan\\University of Arizona, University of Arizona, and Princeton University}
\date{\today}
\maketitle

\begin{abstract}
Many high dimensional classification techniques have been proposed in the literature based on sparse linear discriminant analysis (LDA). To efficiently use them, sparsity of linear classifiers is a prerequisite. However, this might not be readily available in many applications, and rotations of data are required to create the needed sparsity. In this paper, we propose a family of rotations to create the required sparsity. The basic idea is to use the principal components of the sample covariance matrix of the pooled samples and its variants to rotate the data first and to then apply an existing high dimensional classifier. This rotate-and-solve procedure can be combined with any existing classifiers, and is robust against the sparsity level of the true model.  We show that these rotations do create the sparsity needed for high dimensional classifications and provide theoretical understanding why such a rotation works empirically. The effectiveness of the proposed method is demonstrated by a number of simulated and real data examples, and the improvements of our method over some popular high dimensional classification rules are clearly shown.
\end{abstract}

\section{Introduction}
Linear discriminant analysis (LDA) is a useful classical tool for classification.
Consider two $p$-dimensional normal distributions with the same covariance matrix, $N(\bmu_1,\bSigma)$ for class 1 and $N(\bmu_2,\bSigma)$ for class 2. Given a random vector $\bX$ which is from one of these distributions with equal prior probabilities,  a \emph{linear discriminant rule}
\begin{eqnarray}\label{1.1}
\psi_{\bomega,\bnu}(\bX)=I\{(\bX-\bnu)^{\top}\bomega\geq0\}, \quad \bomega,\bnu\in\mathbb{R}^p,
\end{eqnarray}
assigns $\bX$ to class 1 when $\psi_{\bomega,\bnu}(\bX)=1$ and class 2 otherwise. Geometrically, the equation $(\bx-\bnu)^{\top}\bomega=0$ defines an affine space passing through a point $\bnu$ with a normal vector $\bomega$, which is the discriminant boundary of the classification rule.

When $\bmu_1$, $\bmu_2$ and $\bSigma$ are known, the optimal classifier, namely the Fisher linear discriminant rule, is
\begin{eqnarray}\label{1.2}
\psi_F(\bX)=I\{(\bX-\bmu)^{\top}\bSigma^{-1}\bdelta\geq0\},
\end{eqnarray}
where $\bmu=\frac12(\bmu_1+\bmu_2)$, $\bdelta=\bmu_1-\bmu_2$. In practice, these parameters are unknown and replaced by their estimates. Let $\{\bX^{(1)}_i:1\leq i\leq n_1\}$ and $\{\bX^{(2)}_i:1\leq i\leq n_2\}$ be independent and identically distributed (IID) observations from $N(\bmu_1,\bSigma)$ and $N(\bmu_2,\bSigma)$, respectively. In the classical setting with $n_1,n_2\gg p$, $\bmu_1$, $\bmu_2$ and $\bSigma^{-1}$ are usually estimated by sample means $\hat\bmu_1=\bar{\bX}^{(1)}$, $\hat\bmu_2=\bar{\bX}^{(2)}$ and the inverse pooled sample covariance matrix $\hat{\bSigma}^{-1}$. The standard linear discriminant analysis (LDA) uses an empirical version of (\ref{1.2})
\begin{eqnarray}\label{1.3}
\psi_{\hat{F}}(\bX)=I\{(\bX-\hat{\bmu})^{\top}\hat{\bSigma}^{-1}\hat{\bdelta}\geq0\},
\end{eqnarray}
where $\hat\bmu=\frac12(\hat\bmu_1+\hat\bmu_2)$, $\hat\bdelta=\hat\bmu_1-\hat\bmu_2$.

Although the standard LDA has been widely used in applications, it does not work well for high dimensional data when $p$ is comparable to or larger than the sample size. The reason is that, with limited number of observations, it is impossible to estimate too many parameters simultaneously and accurately. In particular, $\hat{\bSigma}$ is singular and not invertible when $n_1+n_2<p-1$. One may use pseudo-inverse $\hat{\bSigma}^-$, but \cite{BickelLevina:2004} showed the LDA performs as poorly as random guessing when $p/(n_1+n_2)\to\infty$. Since the work of \cite{BickelLevina:2004}, a series of LDA-based methods have been proposed for the high dimensional classification problem. The main idea is to find methods which work well when the original classification problem is (nearly) sparse so that $\bmu$ or $\bbeta=\bSigma^{-1}\bdelta$ in the optimal rule (\ref{1.2}) can be well estimated. Ignoring the covariances among the features, \cite{BickelLevina:2004} proposed an independence rule (IR) which outperforms standard LDA in the high dimensional setting. \cite{FanFan:2008} proposed the features annealed independence rule (FAIR) that selects a subset of features before applying the independence rule. In spite of the clear interpretations of the sparsity of the covariance matrix $\bSigma$ and difference of centroids $\bdelta$, in practice, it might be more efficient to find the sparse discriminant affine space directly (see \cite{TrendafilovJolliffe2007,WuETAL:2009,CaiLiu:2011,FanFengTong:2012,MaiZouYuan:2012} among others). Here, a sparse discriminant affine space is an affine space with a sparse normal vector. In particular, \cite{FanFengTong:2012} and \cite{CaiLiu:2011} clearly illustrated the advantages of their direct approaches over IR and FAIR, which over-simplify the problem in many scenarios.

For all aforementioned LDA-based high dimensional classification rules, various explicit sparsity conditions on one or some of $\bSigma$, $\bSigma^{-1}$, $\bdelta$ and $\bbeta$ are crucial to the classification accuracy. For example, IR \citep{BickelLevina:2004} works well only when $\bSigma$ is nearly diagonal; FAIR \citep{FanFan:2008} needs ideally diagonal $\bSigma$ and sparse $\bdelta$; ROAD \citep{FanFengTong:2012} and LPD \citep{CaiLiu:2011} need $\bbeta$ to be sparse to achieve optimal classification. We shall refer to all of these methods as sparse LDA methods. They are efficient when the corresponding sparsity conditions are granted. However, they may not work well when the sparsity conditions are violated. Although these sparse assumptions make sense in some applications, they can be too restrictive in many scenarios (see \cite{HallJinMiller:2009} and reference therein). It is a natural and challenging question how and to what extent we can sparsify a possibly non-sparse problem.

To solve a non-sparse model, a natural idea is to rotate the data to a nearly sparse setting before applying sparse LDA methods. For example, the classification problem can be easily solved by ROAD and LPD if the normal vector of the optimal discriminant affine space, $\bbeta$, is sparse after a rotation. In order to do this, we need an oracle that can rotate the data to such a sparse setting. For the ideal case when $\bbeta$ is known, there are infinitely many orthogonal matrices which can rotate $\bbeta$ to a sparse vector $(||\bbeta||_2,0,...,0)^{\top}$. However, it is not realistic to approximate such rotations before estimating $\bbeta$ itself.  An alternative way might be to make $\bSigma$ diagonal after a rotation, which is related to principal component analysis (PCA). However, such a rotation does not combine the information of the centroids and tends to get wrong directions with small variances, which may actually be crucial for classification.

In this paper, we propose a class of rotations which balance both mean and variance information. Intuitively, both $\bdelta$ and $\bSigma$ should play essential roles in a rotation to make $\bbeta$ sparse. In particular, if $\bSigma$ is spiked \citep{johnstone2001distribution}, its principal components and $\bdelta$ span a linear space, which contains key information on the rotation. Following this intuition, we define $\bSigma^{tot}_{\rho}=\bSigma+\rho\bdelta\bdelta^{\top}$ for $\rho>0$ , whose principal components are determined by the ones of $\bSigma$ as well as $\bdelta$. Consider an orthogonal matrix $\bU_{\rho}$, formed by the eigenvectors of $\bSigma^{tot}_{\rho}$, which diagonalizes $\bSigma^{tot}_{\rho}$. We shall show that $\bU^{\top}_{\rho}\bbeta$ is sparse when the covariance matrix $\bSigma$ is spiked. In other words, the eigenvectors of $\bSigma^{tot}_{\rho}$ are good directions to rotate. Similarly, we can define the empirical version $\hat{\bU}_{\rho}$ which diagonalizes $\hat\bSigma^{tot}_{\rho}=\hat\bSigma+\rho\hat\bdelta\hat\bdelta^{\top}$. The rotation $\hat\bU_{\rho}$ is a reasonably good approximation to $\bU_{\rho}$ when $p\ll n$ \citep{johnstone2009consistency} or $p>n$ with some additional conditions \citep{zou2006sparse,fan2013poet}.  In other words, under some conditions on $\bSigma$, $\hat\bU_{\rho}^{\top}\bbeta$ is nearly sparse, regardless of the sparsity level of the original $\bbeta$. Therefore, we propose to rotate the data by $\hat\bU^{\top}_{\rho}$ first before applying ROAD or LPD, when the sparsity level of $\bbeta$ is unknown. While our original motivation is to make $\bbeta$ sparse by rotation, we find that our procedure is equivariant with respect to orthogonal transformation group $\mathrm{O}(p)$ consisting of all rotations. This feature makes our method robust against the sparsity level of $\bbeta$. The advantage of our method is illustrated by numerous simulated and real data examples.

The rest of our paper is organized as follows. Section 2 introduces a family of ideal rotations and analyzes their theoretical properties. In Section 3, we study a rotate-and-solve procedure for classification. Numerical studies on both simulated and real data are demonstrated in Section 4. All proofs are given in the appendix. Various norms of vectors and matrices appear frequently in the paper. For a vector $\ba$, $||\ba||_p$ denote the standard $\ell_p$-norm. For a matrix $\bA$, $||\bA||$ is the spectral norm.

\section{A family of oracle rotations and their properties}
As mentioned in the introduction, the performance of the sparse LDA methods depend highly on the sparsity of $\bbeta$, which is unknown and hard to verify in practice. High dimensional classifiers will work more efficiently if an oracle rotates the data to a sparse setting before applying sparse LDA methods. If $\bbeta$ is known, we can easily rotate $\bbeta$ to a sparse vector $(||\bbeta||_2,0,...,0)^{\top}$. Of course, it is meaningless to mimic such oracle, which motivates us to find other ideal rotations that can be estimated more easily.

Recall that  the distributions of two classes are $N(\bmu_1,\bSigma)$ for class 1 and $N(\bmu_2,\bSigma)$ for class 2. Let $\bmu=\frac12(\bmu_1+\bmu_2)$, $\bdelta=\bmu_1-\bmu_2$, and $$
    \bSigma^{tot}_{\rho}=\bSigma+\rho\bdelta\bdelta^{\top}, \quad \mbox{ for a given } \rho>0.
$$
Consider an orthogonal matrix $\bU_{\rho}$, formed by the eigenvectors of $ \bSigma^{tot}_{\rho}$, which diagonalizes $\bSigma^{tot}_{\rho}$. For easy presentation, we drop the subscript $\rho$ when its value is fixed or clear in the context. Then, without loss of generality by rearranging columns in $\bU$, we assume that $\bU^{\top}\bSigma^{tot}\bU=\bD$ where $\bD=diag(\eta_1,...,\eta_p)$ is the diagonal matrix, consisting of eigenvalues in descending order.

Let $\{\lambda_j\}_{j=1}^p$ be eigenvalues of $\bSigma$, arranged from the largest to the smallest, and $\{\bxi_j\}_{j=1}^p$ be their corresponding eigenvectors. Note that, for repeated eigenvalues, say $\lambda_r=\lambda_{r+1}=\cdots=\lambda_{s}$, $\{\bxi_j\}_{j=r}^s$ can be chosen as any orthonormal basis of the corresponding eigenspace. \cite{johnstone2001distribution} considered a spiked covariance model, where a few large eigenvalues clearly standing out of the rest.

{\bf Condition 1} (Spiked Covariance Structure): Assume that $\lambda_1\geq\cdots\geq\lambda_k>\lambda_{k+1}=\cdots=\lambda_p$ for some integer $k<p$.

\begin{theorem} \label{thm1}
Under Condition 1, we have $||\bU^{\top}\bbeta||_0\leq k+1$. 
\end{theorem}

Theorem 1 shows the sparsity property of $\bU^{\top}\bbeta$ when $\bSigma$ is spiked and $k+1<p$.  In particular, it implies that
$||\bU^{\top}\bbeta||_1/||\bU^{\top}\bbeta||_2\leq\sqrt{k+1}$ by Cauchy-Schwarz inequality.  The boundedness of the $\ell_0$ or $\ell_1$ norm is crucial for sparse LDA methods such as ROAD and LPD to be efficient. For a vector randomly picked on the unit sphere in $\mathbb{R}^p$, the expectation of its $\ell_1$ norm is of order $\sqrt{p}$. Therefore, both $\ell_0$ and $\ell_1$ norms of $\bbeta$ have been greatly reduced after rotation when $k\ll p$.

The condition of Theorem 1 can still be relaxed somehow while keeping  $||\bU^{\top}\bbeta||_1/||\bU^{\top}\bbeta||_2$ bounded.  This is shown in Theorem 2 below.

{\bf Condition 2} (Quasi-Spiked Covariance Structure): Assume that $\lambda_k\geq\lambda_{k+1}+d$ and $\lambda_{k+1} - \lambda_p \leq \epsilon$ for some integer $k<p$, where $d,\epsilon>0$.

Let $\bW_1$ and $\bW_2$ be two linear spaces spanned by $\{\bxi_j\}_{1\leq j\leq k}$ and $\{\bxi_j\}_{k+1\leq j\leq p}$, respectively. Then, we have $\mathbb{R}^p=\bW_1\oplus\bW_2$ and the mean difference vector $\bdelta$ can be decomposed as $\bdelta=\bdelta_1+\bdelta_2$ with $\bdelta_1\in\bW_1$ and $\bdelta_2\in\bW_2$.

\begin{theorem}\label{thm2}
If $\bdelta\in\bW_1$ and $\lambda_k>\lambda_{k+1}$, then $||\bU^{\top}\bbeta||_0\leq k$ and
$$
    ||\bU^{\top}\bbeta||_1/||\bU^{\top}\bbeta||_2\leq\sqrt{k}.
$$
If $\bdelta\notin\bW_1$ and Condition 2 holds, then
$$
    ||\bU^{\top}\bbeta||_1/||\bU^{\top}\bbeta||_2\leq\sqrt{k+1}+\sqrt{p-k-1}\frac{\lambda_p+\epsilon}{\lambda_p}
    \left (\frac{\epsilon}{\lambda_p} +\sqrt{\frac{\epsilon}{\tilde d-2\epsilon}} \right ),
$$
provided $\epsilon<\tilde{d}/2$, where $\tilde d=d\frac{\rho||\bdelta_2||^2_2}{d+\rho||\bdelta||^2_2}$.
\end{theorem}

Theorem~\ref{thm1} and the first part of Theorem~\ref{thm2} demonstrate that the sparsity can be achieved  after rotation even measured by the strong notion $\ell_0$-norm.  However, the weaker measure of sparsity using $\ell_1$-norm is needed in order to obtain more general results, as shown in the second part of Theorem~\ref{thm2}.

As a direct consequence of Theorem \ref{thm2}, we have the following corollary.
\begin{corollary}
If $\frac{\epsilon}{\lambda_p}=O(\sqrt{k/p})$ and $\frac{\epsilon||\bdelta||^2_2}{d||\bdelta_2||^2_2}=O({k}/{p})$, then $||\bU^{\top}\bbeta||_1/||\bU^{\top}\bbeta||_2=O(\sqrt{k}).$
\end{corollary}

Note that the construction of $\bU$ is independent of $k$, and conclusions of Theorem 2 hold for any $k$ satisfying the technical conditions. Define $d_k=\lambda_{k}-\lambda_{k+1}$, $\epsilon_k=\lambda_{k+1}-\lambda_p$, and $\bW_1^k=\mbox{span}\{\bxi_j\}_{1\leq j\leq k}$, $\bW_2^k=\mbox{span}\{\bxi_j\}_{k+1\leq j\leq p}$, $\bdelta=\bdelta^k_1+\bdelta^k_2$ with $\bdelta^k_m\in\bW^k_m$, $m=1,2$. Let $\tilde d_k=d_k\frac{\rho||\bdelta^k_2||^2_2}{d_k+\rho||\bdelta||^2_2}$. Define $C_k=\sqrt{k+1}+\sqrt{p-k-1}\frac{\lambda_p+\epsilon_k}{\lambda_p}(\frac{\epsilon_k}{\lambda_p} +\sqrt{\frac{\epsilon_k}{\tilde d_k-2\epsilon_k}})$ if $\tilde d_k-2\epsilon_k>0$, and $C_k=\infty$ otherwise. Theorem 2 implies the following corollary.
\begin{corollary}
If $K$ is the least integer such that $\bdelta\in \bW_1^K$, then $||\bU^{\top}\bbeta||_1/||\bU^{\top}\bbeta||_2\leq\min\{C,\sqrt{K}\}$, where $C=\min_{1\leq k< K}\{C_k\}$.
\end{corollary}

Theorems 1 and 2 show that the classification problem is reduced to a sparse one after rotation by $\bU^{\top}$ when the covariance structure is spiked. And the sparsity level of $\bU^{\top}\bbeta$ can be controlled by the spiked covariance structure ($k$ and eigenvalue distribution in Conditions 1 and 2).

Moreover, the procedure is invariant under orthonormal transformations. In other words, the normal vector of the optimal discriminant affine space after rotation, i.e., $\bU^{\top}\bbeta$, is invariant with respect to any rotation. Indeed, when the data are rotated by an arbitrary orthogonal matrix $\bV$, then the new mean vectors and common covariance matrix are $\bV\bmu_1$, $\bV\bmu_2$ and $\bV\bSigma\bV^{\top}$. Since
$$
  \bD=\bU^{\top}\bSigma^{tot}\bU=(\bV\bU)^{\top}\bV\bSigma^{tot}\bV^{\top}(\bV\bU),
$$
the rotation matrix should be $(\bV\bU)^{\top}$, and the rotated normal vector $(\bV\bU)^{\top}\bV\bbeta=\bU^{\top}\bbeta$, which is independent of $\bV$.

\section{A Rotate-and-Solve Procedure}
In this section, we introduce a two-stage rotate-and-solve (RS) procedure for classification. The idea is to mimic the oracle rotations in the previous section and rotate the data such that $\bbeta$ is nearly sparse. Namely, we
first use the orthogonal matrix $\hat{\bU}_{\rho}$, consisting of the eigenvectors of the empirical total covariance $\hat\bSigma^{tot}_{\rho}=\hat\bSigma+\rho\hat\bdelta\hat\bdelta^{\top}$ to rotate the data and then apply sparse LDA methods such as ROAD and LPD to the rotated data.

Let $\hat{\bmu}_1$ and $\hat{\bmu}_2$ be the sample mean vectors of classes 1 and 2 respectively.  Set
\[
    \hat\bmu=(\hat\bmu_1+\hat\bmu_2)/2, \quad \mbox{and} \quad \hat\bdelta=\hat\bmu_1-\hat\bmu_2.
\]
Similarly, let $\hat\bSigma^{(1)}$ and $\hat\bSigma^{(2)}$ be their sample covariance matrices and
\[
    \hat\bSigma=\frac{1}{n_1+n_2}(n_1\hat\bSigma^{(1)}+ n_2 \hat\bSigma^{(2)})
\]
be the pooled sample covariance matrix.  The degree of freedom can be adjusted, but the version of the maximum likelihood estimate (MLE) is used here to facilitate the expression in Remark 1 below.  We then estimate $\bSigma^{tot}_{\rho}$ by
\[
    \hat\bSigma^{tot}_{\rho}=\hat\bSigma+\rho\hat\bdelta\hat\bdelta^{\top},
\]
whose dependence on $\rho$ will be temporarily dropped for easy presentation. Perform singular-value decomposition
\begin{eqnarray}\label{t2.4}
\hat\bU^{\top}\hat\bSigma^{tot}\hat\bU=\hat\bD,
\end{eqnarray}
where $\hat\bD= \mathrm{diag}(\hat\eta_1,...,\hat\eta_p)$ is the diagonal matrix with sorted eigenvalues.

The two-stage rotate-and-solve procedure can be implemented as follows.
\begin{itemize}
\item []
\textbf{Stage one}: Calculate $\hat\bU$ and rotate the data to get $\{\hat{\bU}^{\top} \bX_i^{(m)}\}_{i=1}^{n_m}$
 for $m=1$ and $2$.

\item [] \textbf{Stage two}: apply ROAD, LPD or other sparse LDA methods to the rotated data $\cX \hat\bU$ to get a prediction rule. 
\end{itemize}

{\bf Remark 1}: Define
\[ \bar\bX=\frac{1}{n_1+n_2}(n_1\bar\bX^{(1)}+n_2\bar\bX^{(2)}),\]
\[\hat\bSigma^{tot}_{sample}=\frac{1}{n_1+n_2}\sum_{m=1}^2\sum_{i=1}^{n_m}(\bX^{(k)}_i-\bar\bX)(\bX^{(k)}_i-\bar\bX)^{\top}\]
which is the sample total covariance (ignoring the classes). It is straightforward to see $\hat\bSigma^{tot}_{sample}=\hat\bSigma+\frac{n_1n_2}{(n_1+n_2)^2}\hat\bdelta\hat\bdelta^{\top}$.

When $p\ll n=n_1+n_2$, $\hat\bU$ and $\bU$ are similar when the eigenvalues are separated from each other, and hence $\hat\bU^{\top}\bbeta$ is similar to $\bU^{\top}\bbeta$. The property of $\hat\bU^{\top}\bbeta$ is much more complicated when $p\sim n$ or $p\gg n$.  In this case, it is hard to guarantee all estimated eigenvectors are close to the true ones.  However, the eigenvectors that correspond to spiked eigenvalues can be consistently estimated.  See for example
\cite{zou2006sparse,karoui2008operator,johnstone2009consistency, agarwal2012noisy, fan2013poet, shen2013surprising}.
As these eigenvectors point at most important directions, the consistent estimation of these directions ensures the correct rotations in these important directions.  This explains our empirical results that the RS procedure performs very well compared to several state-of-the-art methods, even when $p\gg n$.

To understand better the mathematics behind the excellent performance of RS procedure, the classification error of the idealized Fisher classifier depends on $\gamma \equiv \bdelta^{\top} \bSigma^{-1} \bdelta$.  Let $\bU_1$ be a $(k+1)\times p$ matrix, consisting of the eigenvectors of  $\bSigma^{tot}$ that correspond to the largest $k+1$ eigenvalues $\{\eta_j\}_{j=1}^{k+1}$.  If we restrict the information to the first $k+1$ dimensions of the rotated data $\bU_1^{\top} \bX|_m \sim N(\bU_1^{\top} \bmu_m, \bU_1^{\top} \bSigma \bU_1)$, $m=1,2$, then the classification error depends on
$$
   \gamma_1 \equiv (\bU_1^{\top} \bdelta)^{T} (\bU_1^{\top} \bSigma \bU_1)^{-1} (\bU_1^{\top} \bdelta).
$$
Clearly, $\gamma_1 \leq \gamma$.  How much is the information loss when $\{\eta_j\}_{j=1}^{k+1}$ are spiked?  Under Conditions in Theorem 1, there is no information loss if the first $k+1$ most important features are used.  Furthermore, the cited literatures above give the conditions under which $\bU_1$ can be consistently estimated.

The above argument is based on the fact that $\bU_1^{\top} \bdelta$ preserves the energy of $\bdelta$.  The result holds more generally for the covariance matrix $\bSigma$ admitting spiked eigenvalues, including covariance matrices derived from approximate factor models \citep{fan2013poet} or admitting low rank plus sparse matrix decomposition \citep{agarwal2012noisy}.  Recall that
$\bSigma = \sum_{i=1}^p \lambda_i \bxi_i \bxi_i^{\top}$ with $\bxi_i$ being the eigenvector of $\bSigma$.  Let $\lambda_i(\bB)$ be the $i^{th}$ largest eigenvalue of a symmetric matrix $\bB$.

\begin{theorem}\label{thm3}
If $\lambda_{k+1} (\sum_{i=1}^k \lambda_i \bxi_i \bxi_i^{\top} + \rho \bdelta \bdelta^{\top}) > a \lambda_{k+1}$ for some $a > 2$, then $||\bU_1 \bdelta ||_2 \geq \frac{a-2}{a-1} \|\bdelta \|_2$ and $\gamma_1 \geq \frac{(a-2)^2}{(a-1)^2 \lambda_1} \|\bdelta \|^2_2$.
\end{theorem}

The condition of Theorem 3 holds relatively easily.  We can take $k = 0$ when $\rho \|\bdelta\|^2_2 \geq a \| \bSigma \|^2$.  This holds easily by taking a sufficiently large $\rho$.

Note that $\gamma \leq \lambda_p^{-1} \|\bdelta\|^2_2$ and $\gamma$ is usually significantly smaller than this upper bound.  Therefore, when $\lambda_1/\lambda_p$ is bounded, the loss of information by using rotated data is limited. Yet, we reduce significantly the noise accumulation in classification \citep{FanFan:2008}. As noted above, the rotation $\bU_1$ can be consistently estimated by regularization.  These together provide theoretical endorsement of the advantages of using rotation.

{\bf Remark 2}: (Dimensionality reduction) When $p>n$, $\hat\bU$ is not unique since $\hat\bSigma^{tot}$ is singular. (The null space of $\hat\bSigma^{tot}$ is large and we can choose arbitrary basis of the null space as the columns of $\hat\bU$.) 
Since the last $p-n$ columns in $\hat\bU$ are arbitrary and can not be controlled, we define $\tilde\bU$ as the first $n$ columns (or even fewer) of $\hat\bU$ and conduct classification on the rotated data $\{\tilde\bU^{\top} \bX_i^{(m)}\}_{i=1}^{n_m}$ for $m=1$ and $2$.  From the theoretical analysis in the last section, we see that, under ideal conditions, $\bU^{\top}\bbeta$ is sparse with non-vanishing part concentrated on the first $k+1$ components. This implies that only first $k+1$ columns of the rotated data are useful to estimate $\bU^{\top}\bbeta$, which motivates us to use $\tilde \bU$ instead of $\hat\bU$ as a practical approach with reduced dimensionality. Theorem~\ref{thm3} further shows that the loss of classification power due to this dimensionality reduction is limited.
Let $\tilde\psi$ be a classification rule constructed by some (fixed) sparse LDA method based on $\tilde{\mathcal{X}}=\mathcal{X}\tilde\bU$. It is straightforward to see that $\tilde\psi$ is equivariant.

{\bf Remark 3} (Computation of transform)  When $p > n$, the computation of $\tilde{\bU}$ can be performed as follows.  First of all, $\hat\bSigma^{tot}$ can be written as $\bY^{\top}\bY $ for a given $ (n+1)\times p$ matrix $\bY$ (suitable scaling of centered observations and sample mean).  Note that  $\bY^{\top} \bY$ and $\bY \bY^{\top}$ have the same non-vanishing eigenvalues. Let $\tilde \bU = \bY^{\top} \hat \bV$, where $\hat{\bV}$ is the orthogonal matrix consisting of eigenvectors of non-vanishing eigenvalues of the $(n+1)\times (n+1)$ matrix $ \bY \bY^{\top}$.  Then, the columns of $\tilde \bU$ contain the eigenvectors of nonvanishing eigenvalues of $\bY^{\top} \bY$ and are orthogonal.  In other words, $\tilde \bU$ can be used to transform the data. The reduction of computation cost is significant when $p \gg n$, since the singular value decomposition of $\bY \bY^{\top}$ is much faster.

{\bf Remark 4} (Sensitivity of $\rho$). Our empirical studies show that the rotate-and-solve procedure is not sensitive to $\rho$ in a broad range. For a large range of choices of $\rho$, the classification errors are significantly improved over the existing LDA algorithms, as will be shown by our numerical experiments in the next section. Ideally, $\rho$ can be estimated using data-adaptive methods such as cross-validation. However, cross-validation on $\rho$ may be computationally intractable for high dimensional data where $p$ is huge. As noted from Remark 3, we may use $\tilde\bU$ to rotate the data which reduces the dimension from $p$ to $n$. Thus cross-validation on $\rho$ is more tractable using the modified rotate-and-solve procedure, and the classification quality can be noticeably improved as to be shown by our numerical experiments.

\section{Numerical Studies}
In this section, we compare the rotate-and-solve (RS) procedure with a number of popular LDA-based methods including standard LDA (\ref{1.3}) (using Moore-Penrose pseudoinverse when $\hat\bSigma$ is singular), IR, nearest shrunken centroids (NSC) \citep{tibshirani2002diagnosis}, ROAD and LPD, via simulation and real data examples. For the RS procedure, two variants RS-ROAD and RS-LPD are included. For simulated examples using the toy models, we also consider the oracle RS methods (O-RS-ROAD and O-RS-LPD) where the oracle rotation shown in Section 2 are used to rotate the data. Moreover, the oracle Fisher's rule (\ref{1.2}) is used as a benchmark method. In all RS-related methods, the parameter $\rho$ is fixed to $\frac12$ unless explicitly defined. The same number of observations are generated for both classes for all simulated data in Section \ref{Subsec:ToyModels}, i.e. $n_1=n_2$. All simulation settings have been repeated 100 times unless noted otherwise.

\subsection{Simulated Data}\label{Subsec:ToyModels}

\subsubsection{Toy Models}

We begin with several toy models with relatively small $n$ and $p$ to illustrate the performance of the RS procedures versus aforementioned LDA methods. We consider the following three toy models:
\begin{itemize}
\item {\bf Toy Model 1}. $\bSigma=\bI_p$; $\bmu_1=\bzero_p$ and $\bmu_2=a_1\bone_p$.

\item {\bf Toy Model 2}. $\bSigma=(\sigma_{i,j})$ with $\sigma_{i,i}=1$ and $\sigma_{i,j}=0.5$ for $i\ne j$; $\bmu_1=\bzero_p$ and $\bmu_2=(a_2\bone_\ell^{\top},\bzero^{\top}_{p-\ell})^{\top}$, where $\ell=5$.

\item {\bf Toy Model 3}. The setting is the same as 2 except  $\ell=p/2$, $\bmu_2=(a_3\bone_\ell^{\top},\bzero^{\top}_{p-\ell})^{\top}$.
\end{itemize}
The values of $a_1$, $a_2$ and $a_3$ in each of the toy models are chosen such that the expected classification errors of the oracle Fisher's rule (\ref{1.2}) are 1\%, 5\% and 10\%. For each model, we take $p=50$ and $n_1=20$ or $30$. The same number of observations have been collected independently as the testing set.

\begin{figure}[htp]
\centering
    \includegraphics[width=6.5in]{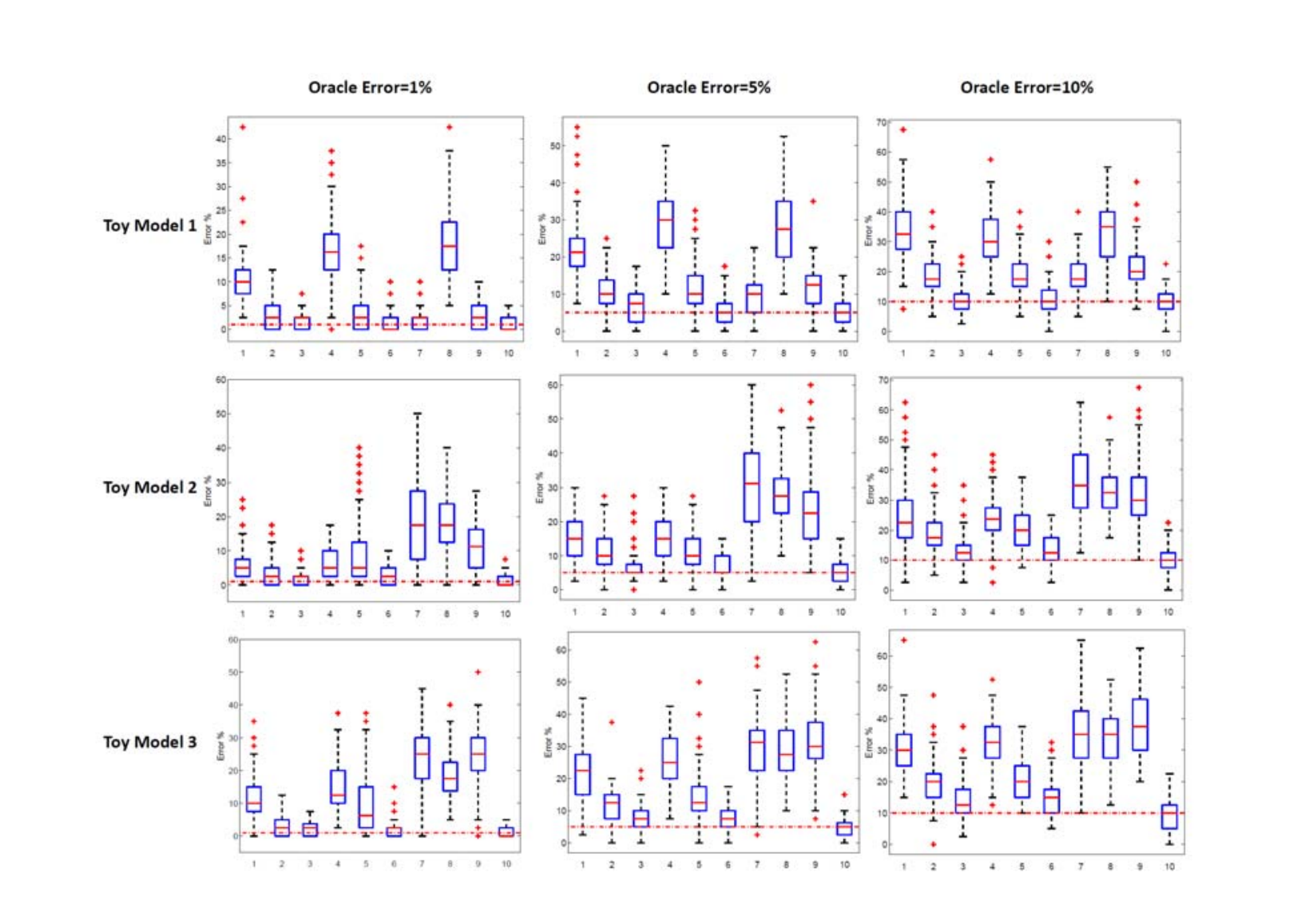}
\caption{{Simulation results for the three toy models with $n_1=20$ and $p=50$. 1=ROAD, 2=RS-ROAD, 3=O-RS-ROAD, 4=LPD, 5=RS-LPD, 6=O-RS-LPD, 7=IR, 8=Standard LDA, 9=NSC, 10=Oracle.}}\label{Fig:ToyModels:n20}
\end{figure}

\begin{figure}[htp]
\centering
    \includegraphics[width=6.5in]{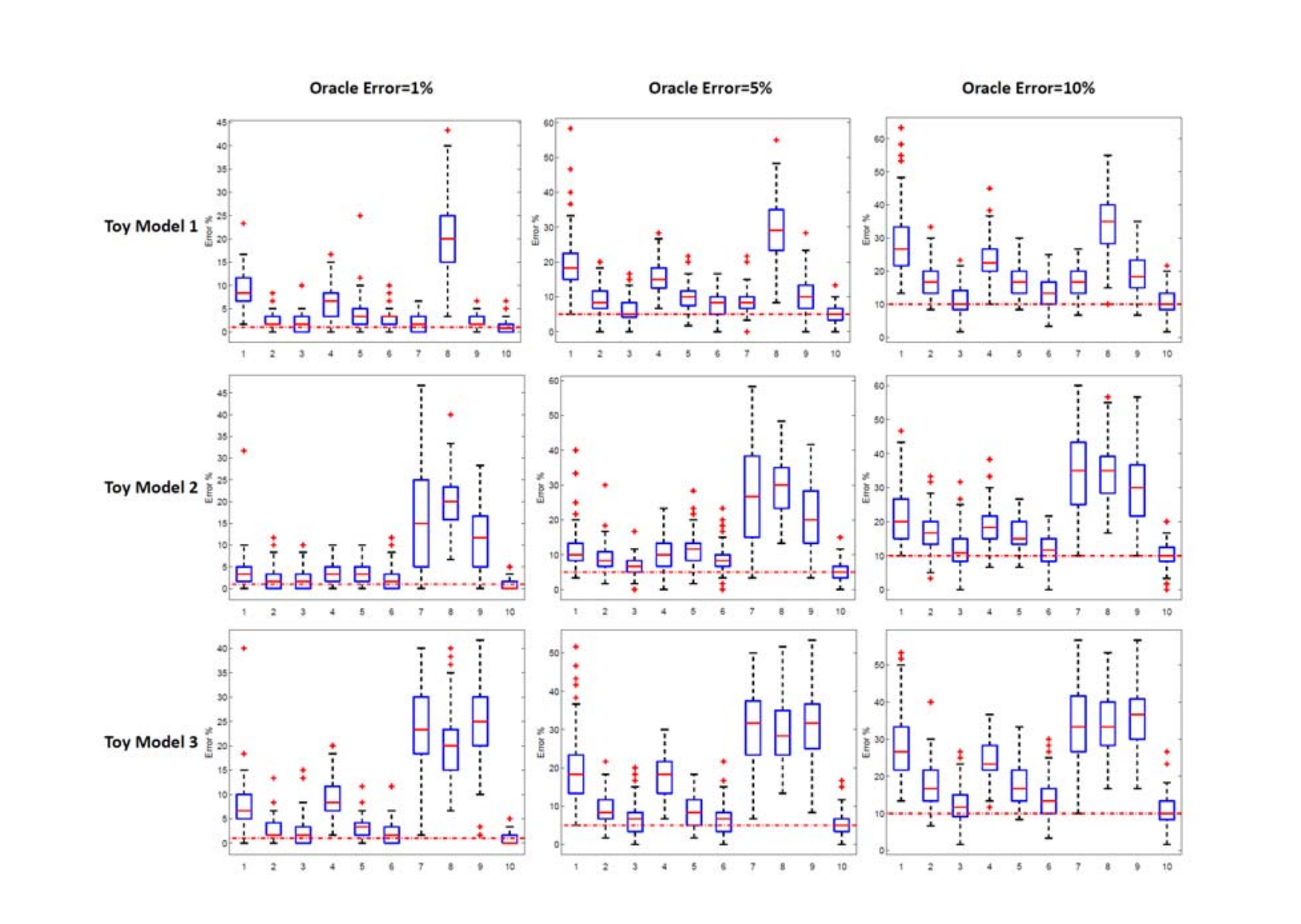}
\caption{{Simulation results for the three toy models with $n_1=30$ and $p=50$. 1=ROAD, 2=RS-ROAD, 3=O-RS-ROAD, 4=LPD, 5=RS-LPD, 6=O-RS-LPD, 7=IR, 8=Standard LDA, 9=NSC, 10=Oracle.}}\label{Fig:ToyModels:n30}
\end{figure}

We apply IR, Standard LDA, NSC, ROAD, LPD, RS-ROAD, RS-LPD, O-RS-ROAD and O-RS-LPD to 100 replicates of every simulation scenario. Simulation results are presented in Figure \ref{Fig:ToyModels:n20} (for $n_1=20$) and Figure \ref{Fig:ToyModels:n30} (for $n_1=30$). The Oracle rule always performs best and gives a benchmark for other methods. The O-RS methods perform very well and are comparable with oracle rule. For toy model 1, the features are independent, so IR performs best besides the oracle rule. But RS methods are comparable with IR. For model 2, the true $\bbeta$ is nearly sparse. Therefore, ROAD and LPD perform well but RS methods still improve their performance. For model 3, neither the covariance matrix nor true $\bbeta$ is sparse. RS methods work significantly better than their competitors. We observe that the RS methods are uniformly good in all the three models. 

\begin{figure}[htp]
\centering
    \includegraphics[width=2.0in]{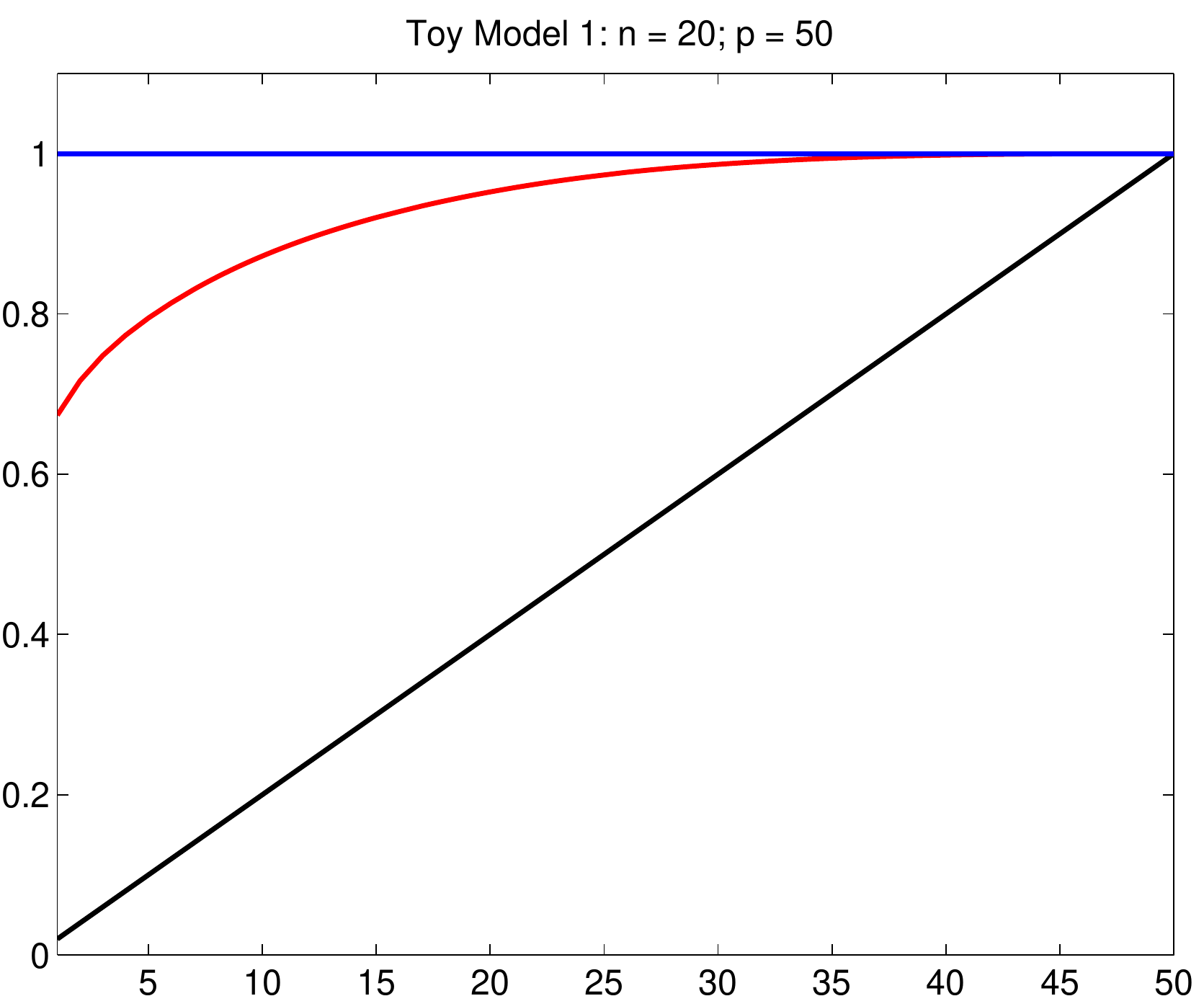}
    \includegraphics[width=2.0in]{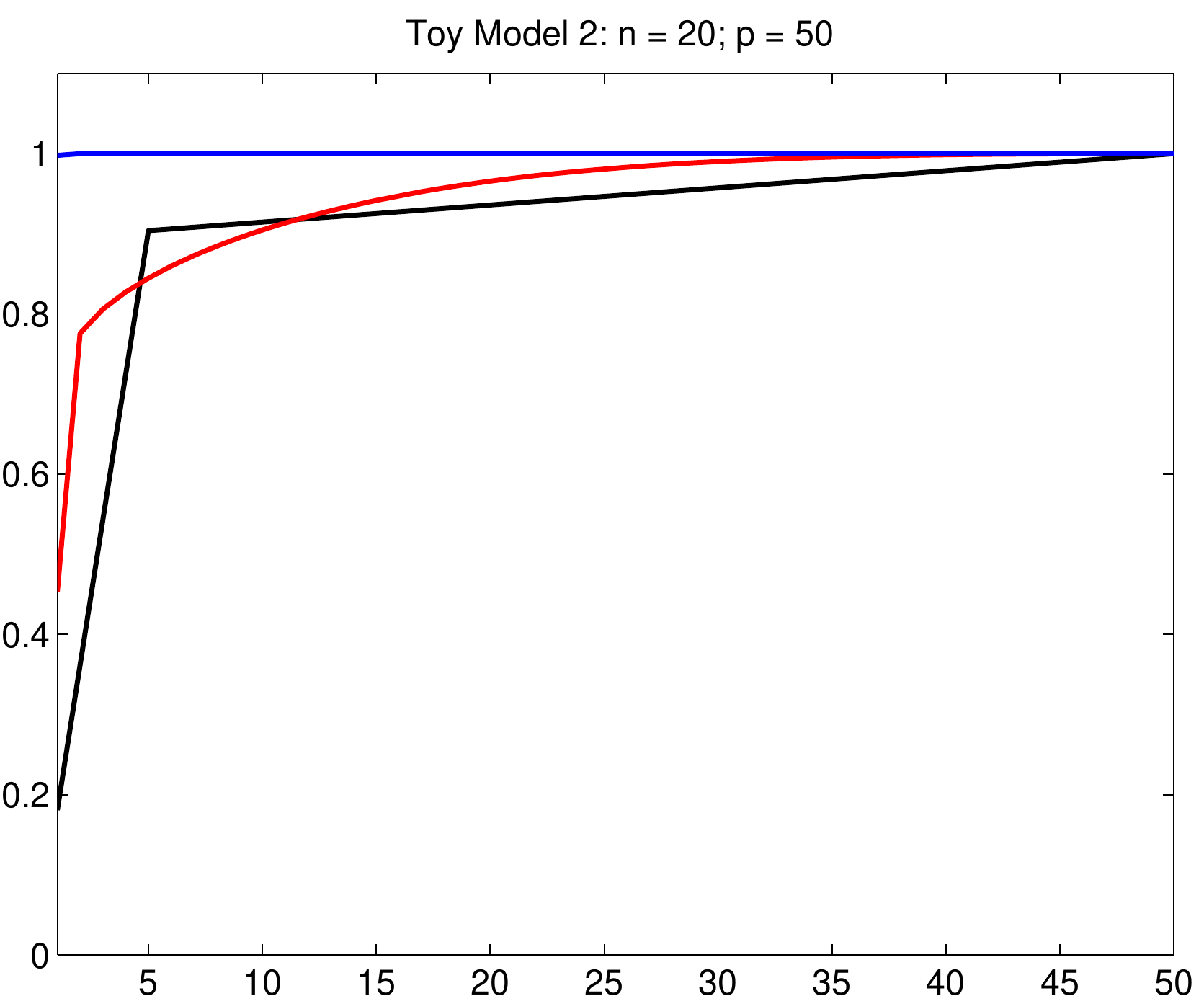}
    \includegraphics[width=2.0in]{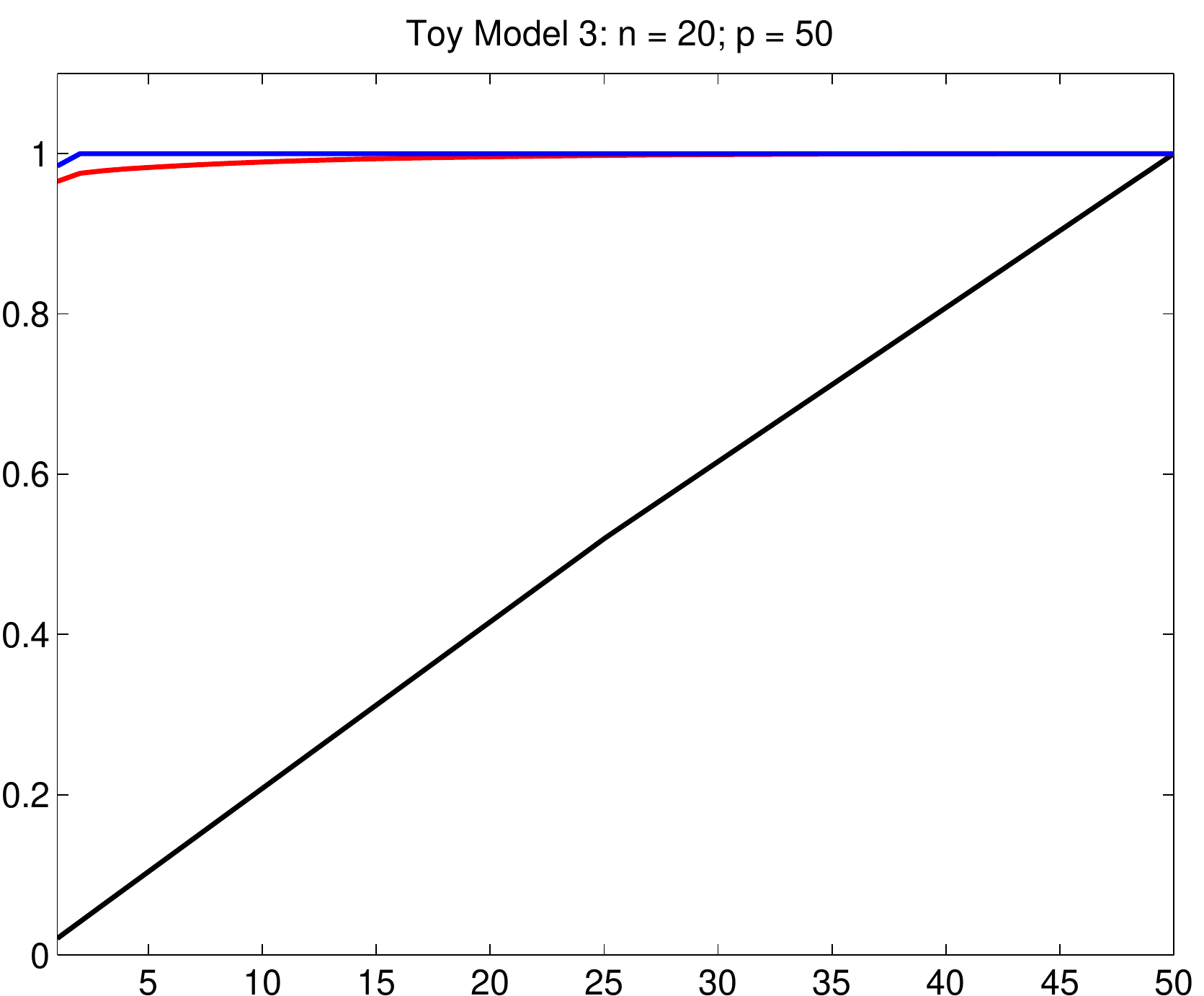}\\
    \includegraphics[width=2.0in]{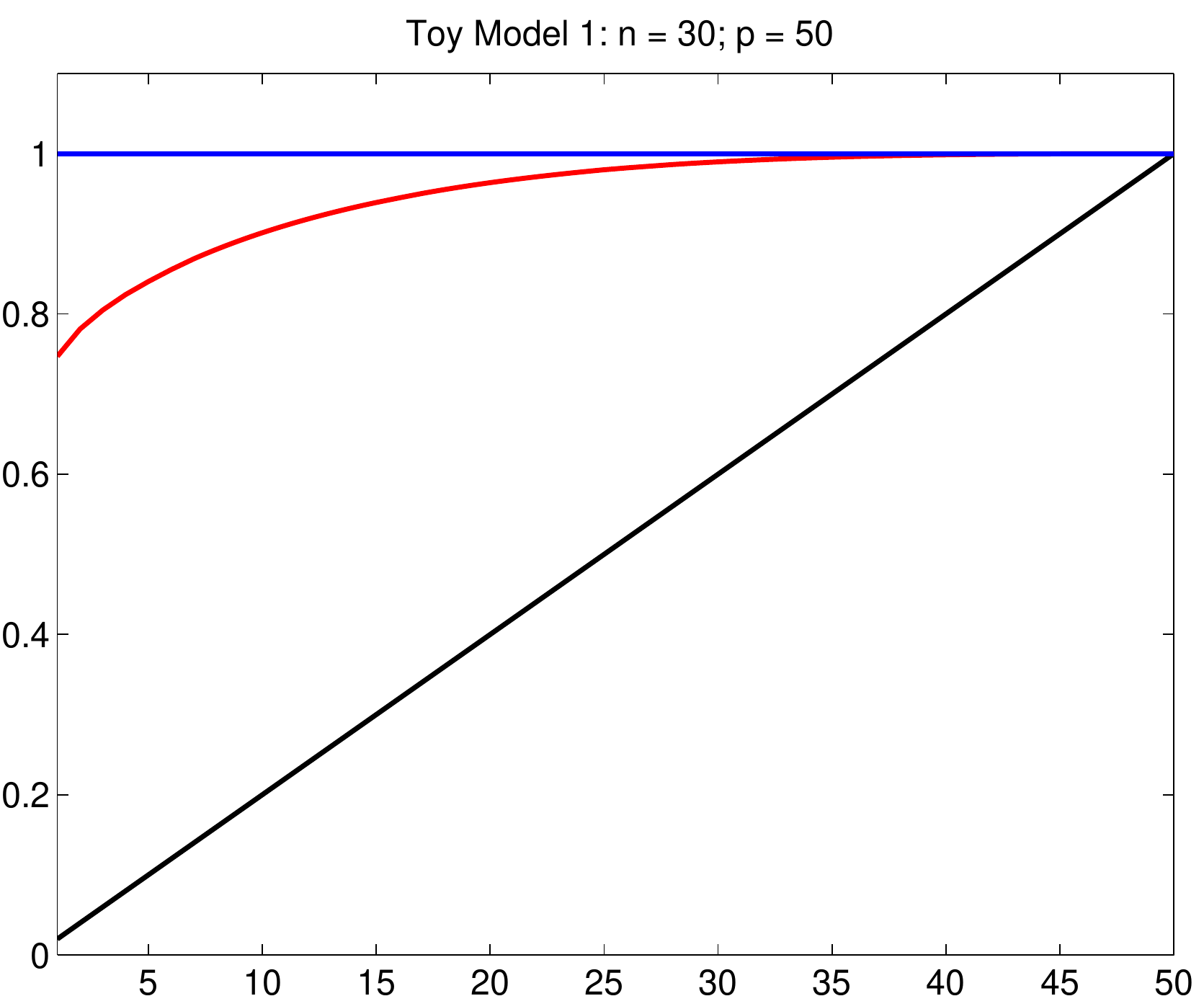}
    \includegraphics[width=2.0in]{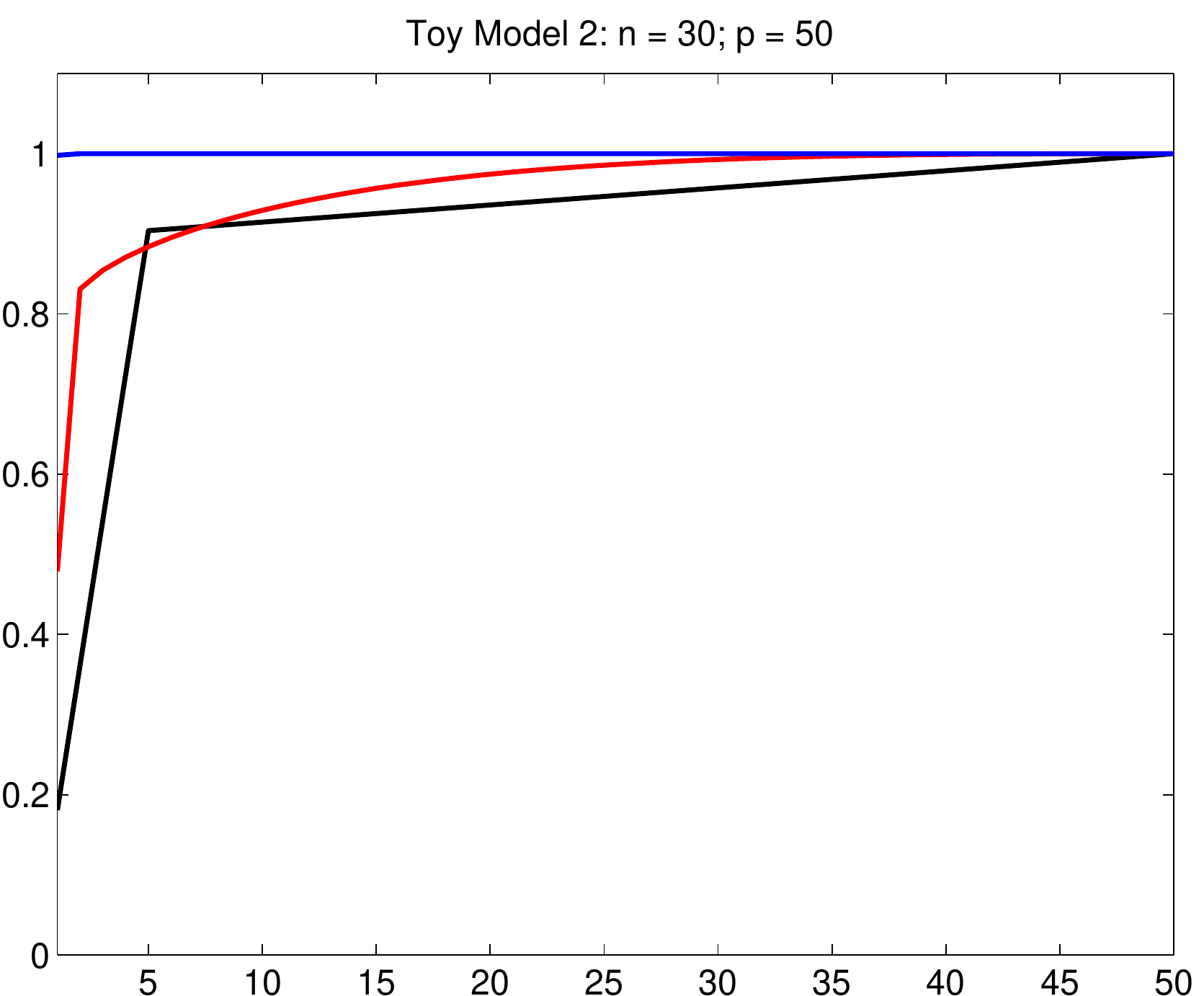}
    \includegraphics[width=2.0in]{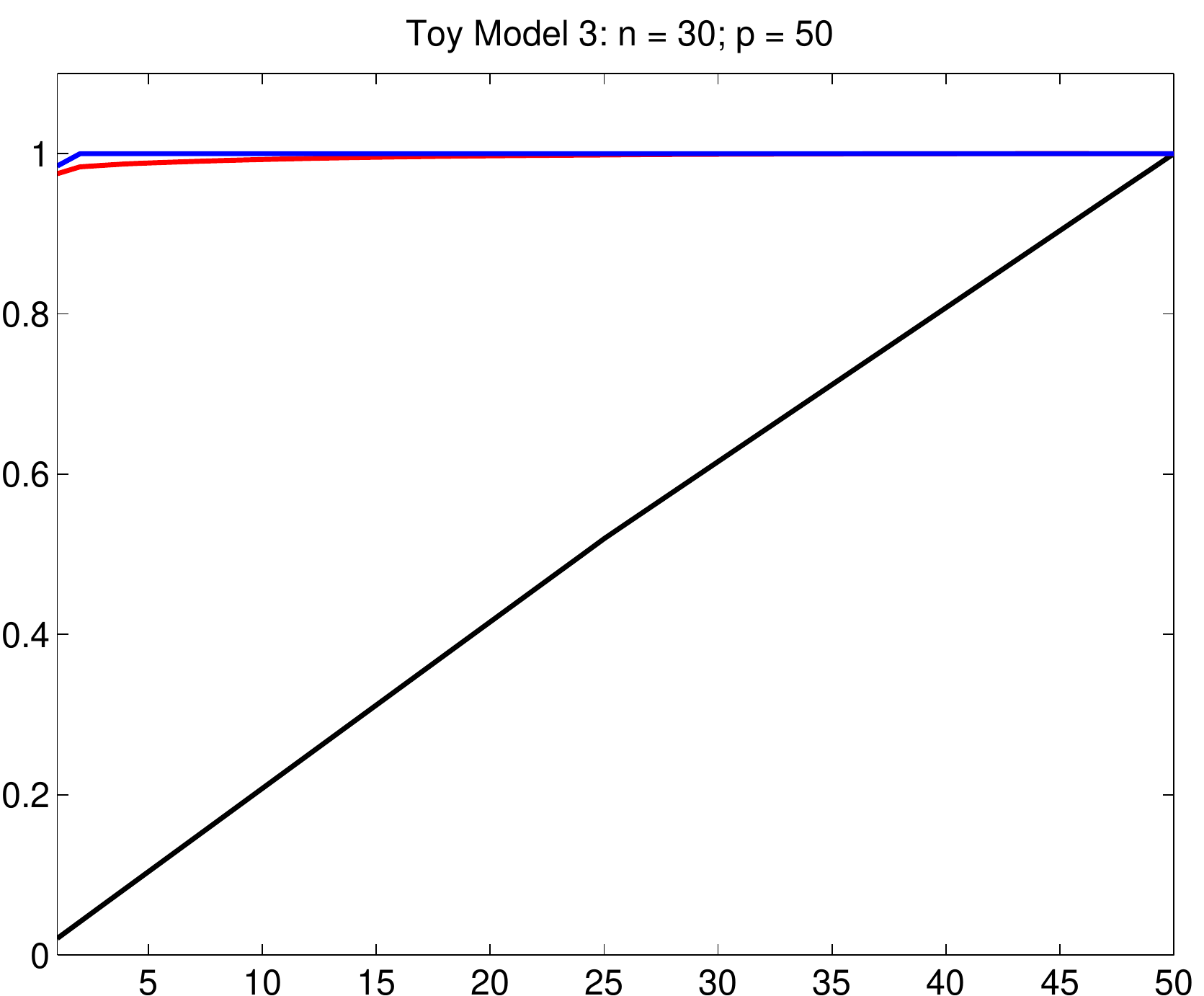}
\caption{Sparsity levels of $\bbeta$ before (black) and after rotation using $\bU$ (blue) and $\hat \bU$ (red).
The top row corresponds to the case $n_1=20$, $p=50$; while the bottom row corresponds to $n_1=30$, $p=50$.}\label{Fig:SparsityAfterRotation}
\end{figure}

To see why RS methods outperform their direct sparse competitors, we plot the percentages of sum squares of the first several largest components of true $\bbeta$ before and after rotation. For a rotation $R=\bU$ or $\hat\bU$, define $\bbeta^{R}=R^{\top}\bbeta$. Denote by $|\beta|_{(1)}, \cdots, |\beta|_{(p)}$ and
$|\beta^{R}|_{(1)}, \cdots, |\beta^{R}|_{(p)}$ the reversed order statistics (from largest to smallest) of $\{|\beta_j|\}_{j=1}^p$ and $\{|\beta^{R}_j|\}_{j=1}^p$, respectively.  For each
setting, we plot $\sum_{i=1}^k|\beta|^2_{(i)}/||\bbeta||^2_2$, $\sum_{i=1}^k|\beta^{\bU}|^2_{(i)}/||\bbeta^{\bU}||^2_2$ and $\frac{1}{100}\sum_{j=1}^{100}\sum_{i=1}^k|\beta^{\hat\bU_{j}}|^2_{(i)}/||\bbeta^{\hat\bU_{j}}||^2_2$ for $k=1$,..., $p$, where $\hat\bU_{j}$ is the rotation matrix for $j$th replicate and $\bU$ is the oracle rotation matrix. In Figure \ref{Fig:SparsityAfterRotation}, we see, after rotation, $\bbeta$ is more concentrated in its largest components. $\bU^{\top}\bbeta$ is extremely sparse, and $\hat\bU^{\top}\bbeta$ is sparser than the original $\bbeta$. Obviously, ROAD/LDP is more efficient after the rotation. 

\vspace{-0.2cm}

\subsubsection{More Simulations}
\vspace{-0.2cm}

\begin{figure}[htp]
\centering
    \includegraphics[width=2in]{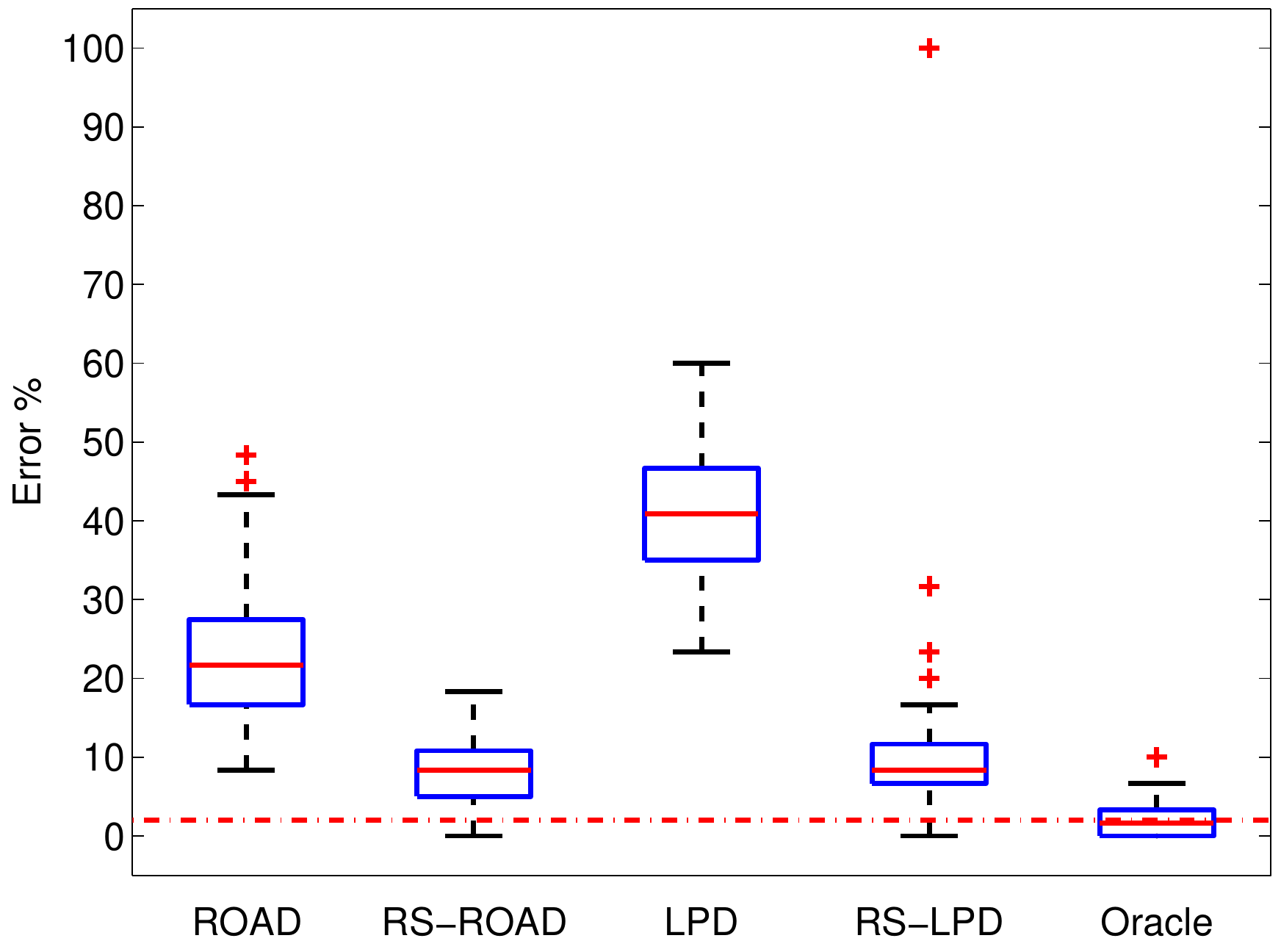}
    \includegraphics[width=2in]{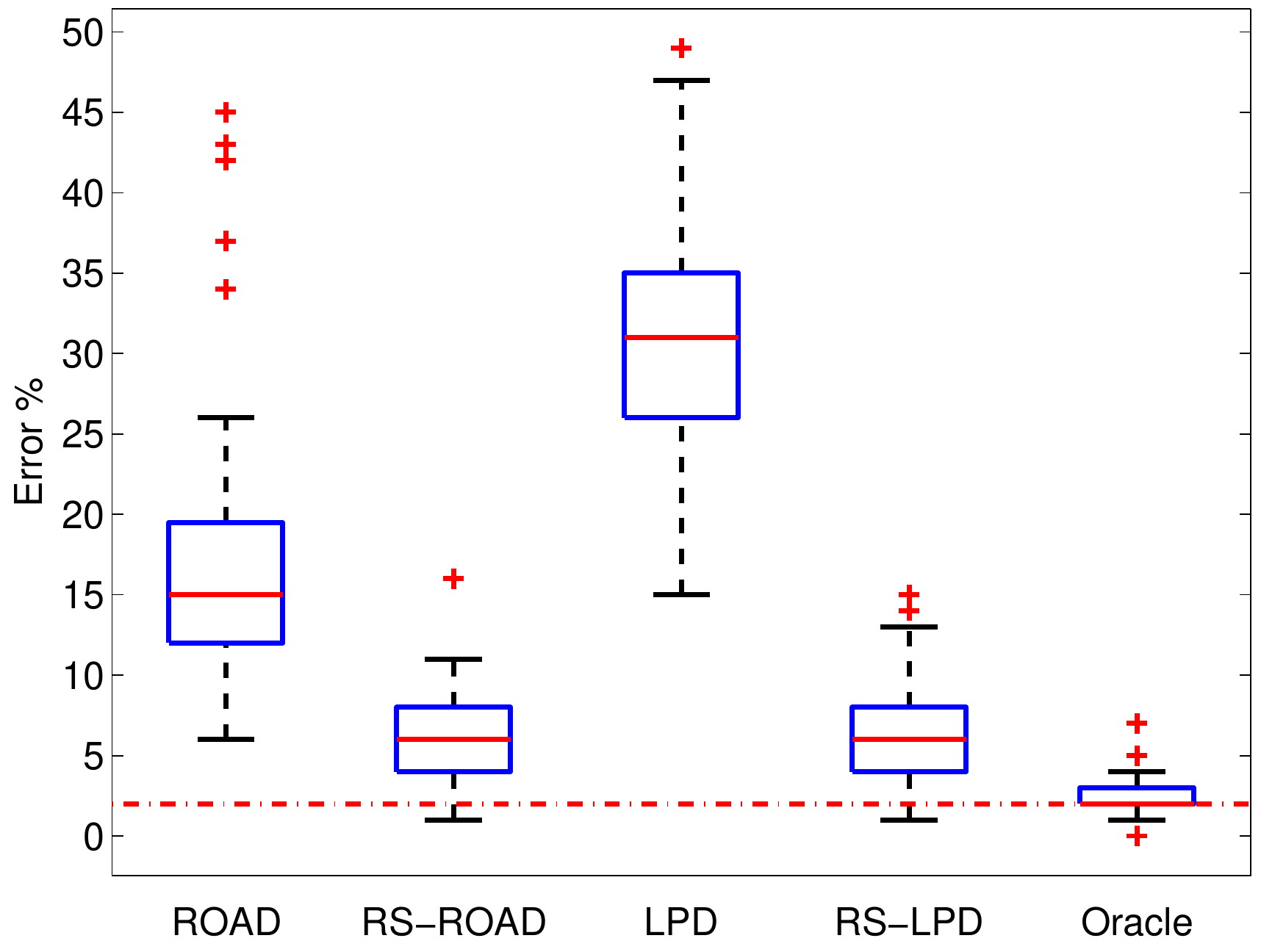}
    \includegraphics[width=2in]{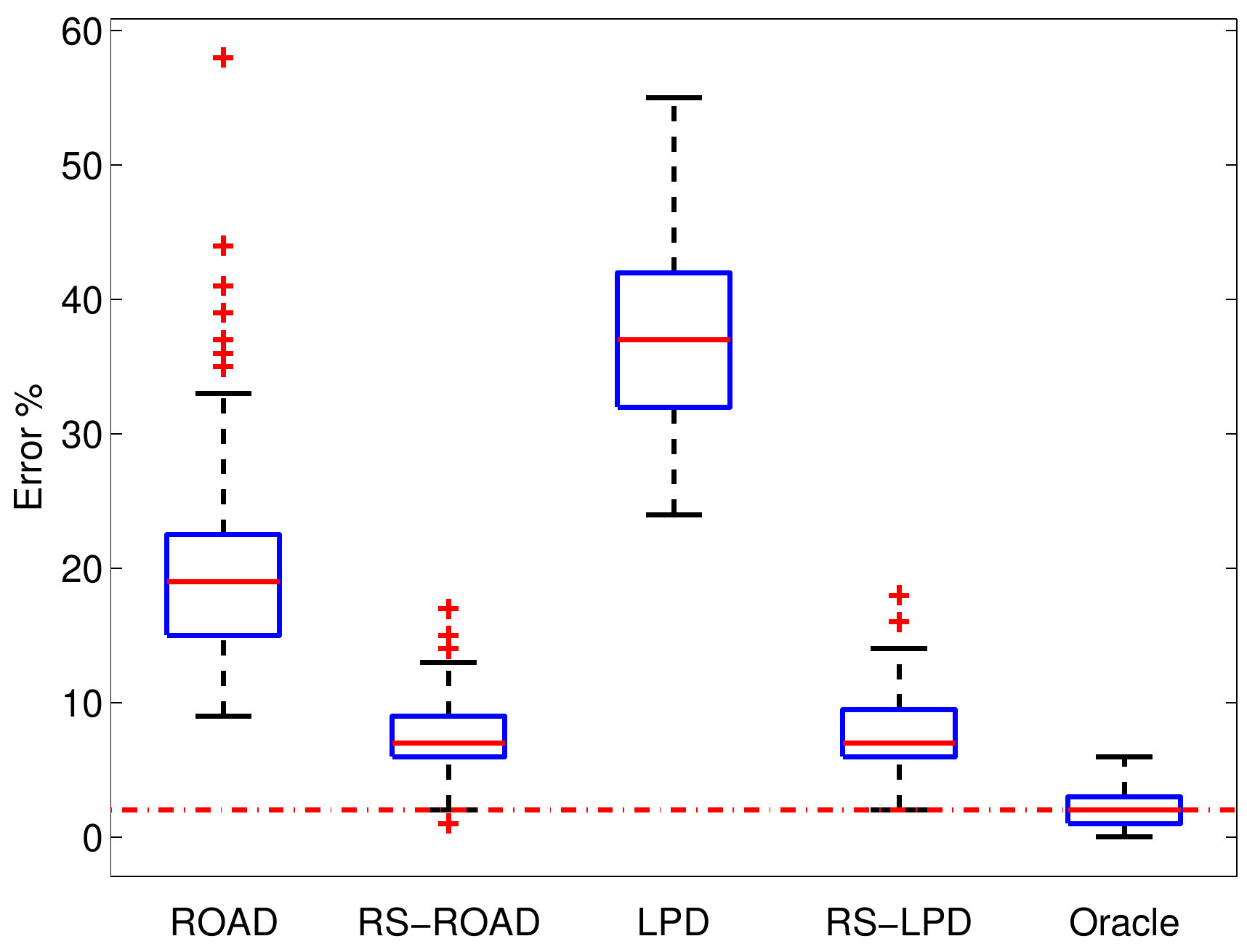}
    \includegraphics[width=2in]{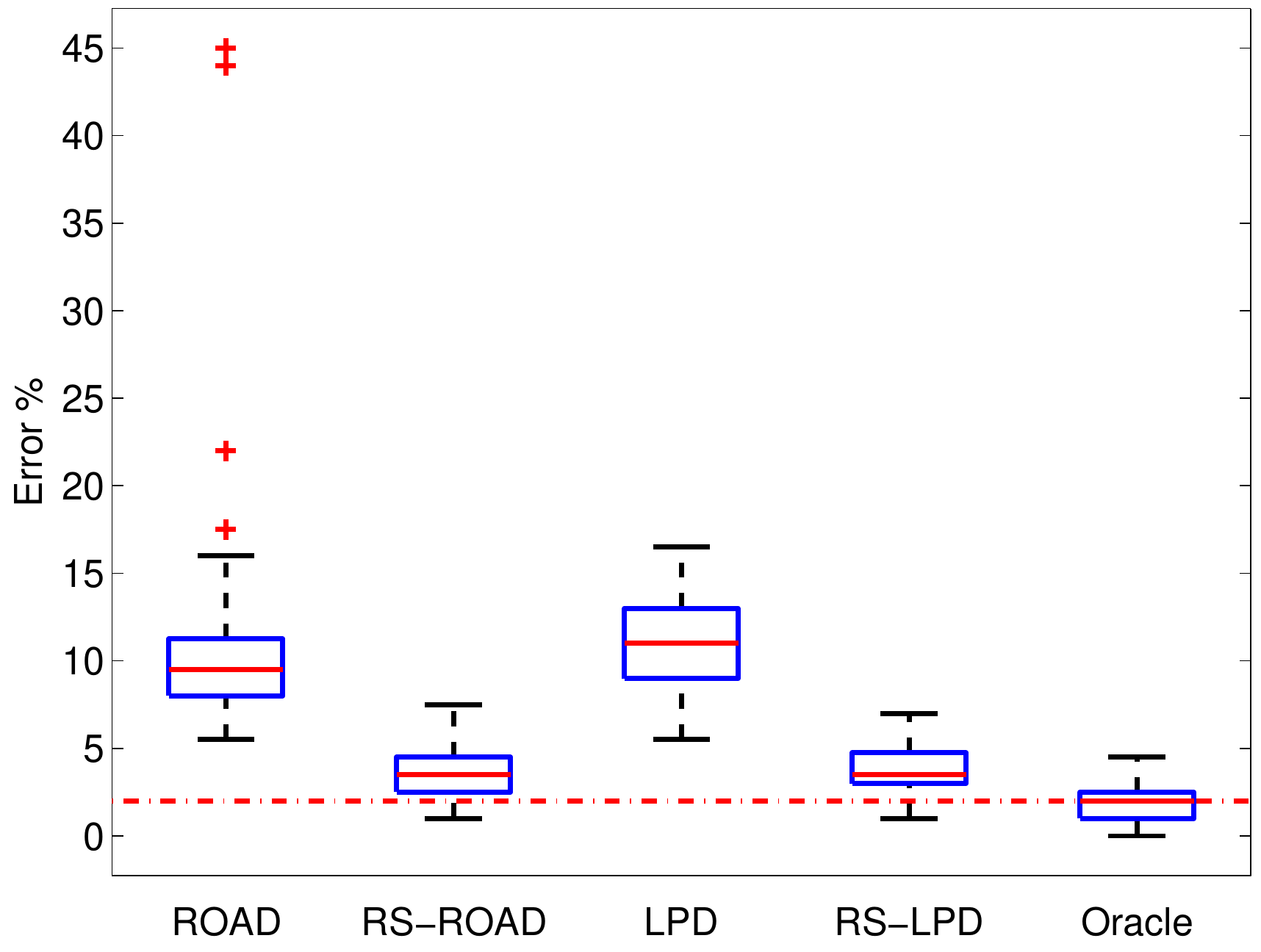}
    \includegraphics[width=2in]{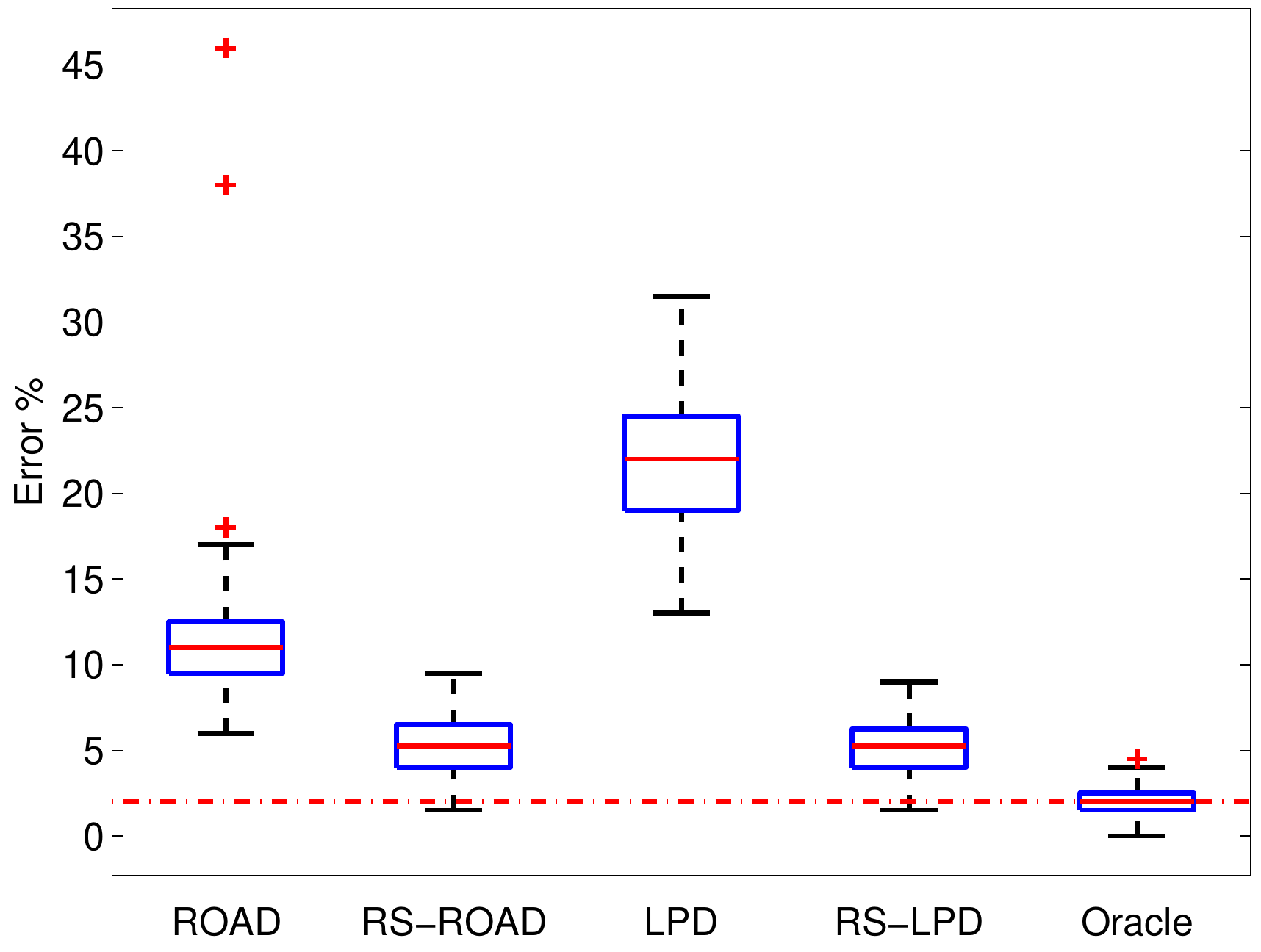}
\caption{Simulation results of Model 1: the boxplots from left to right correspond to the cases $(n_1,p)=(30,
200)$, $(50, 200)$, $(50, 400)$, $(100, 200)$ and $(100, 400)$.}\label{Fig:SecondSetOfModels:1}
\end{figure}

\begin{figure}[htp]
\centering
    \includegraphics[width=2in]{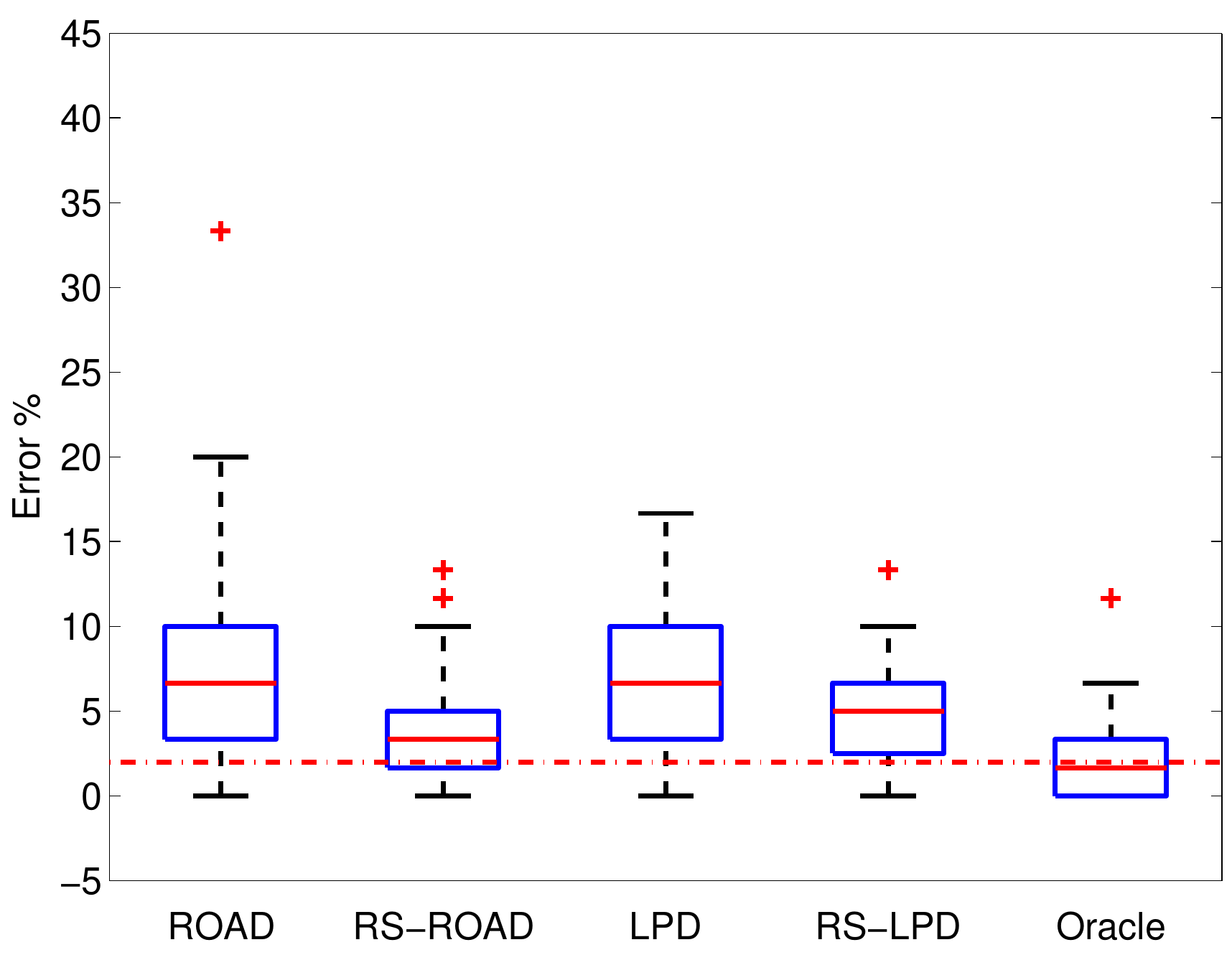}
    \includegraphics[width=2in]{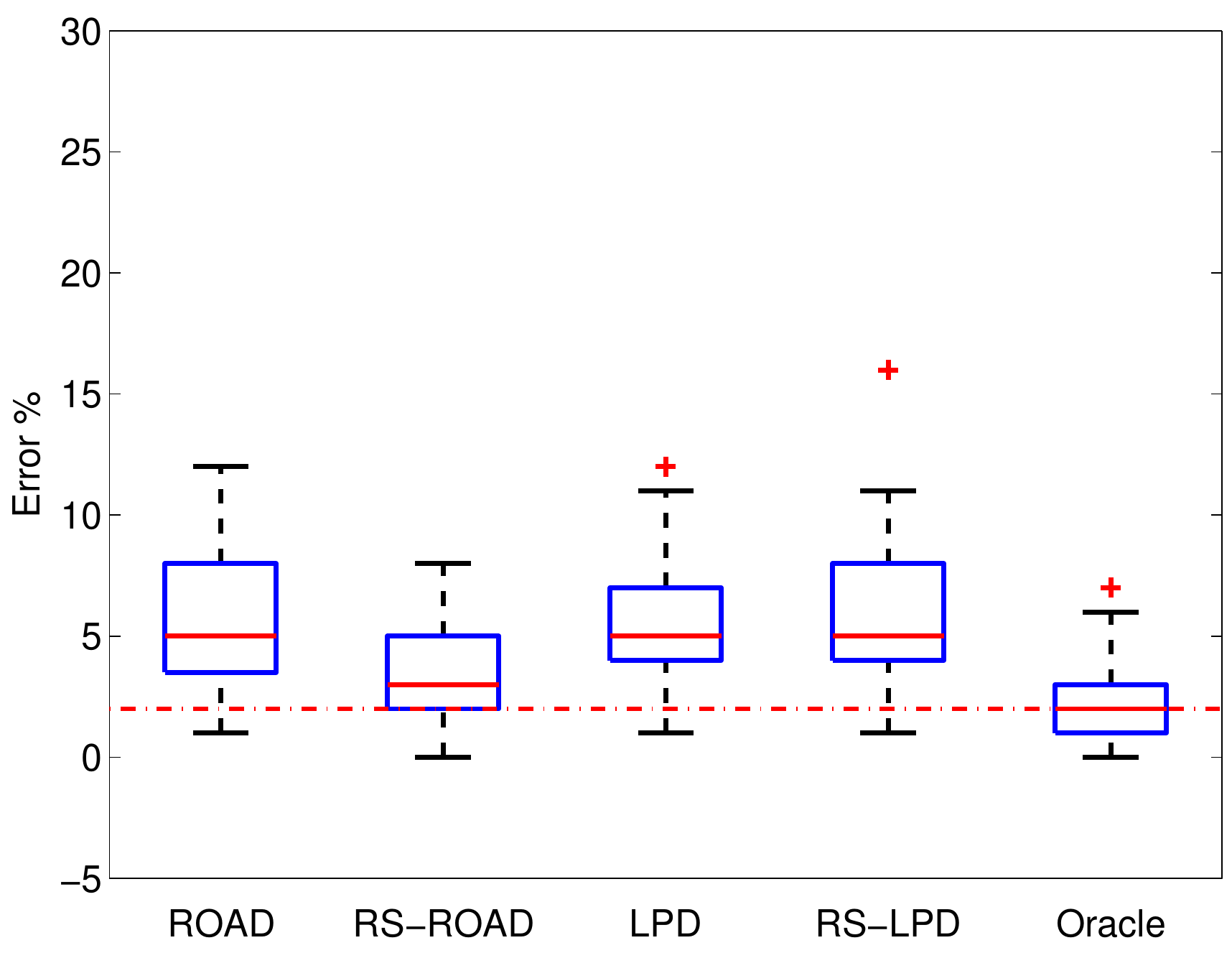}
    \includegraphics[width=2in]{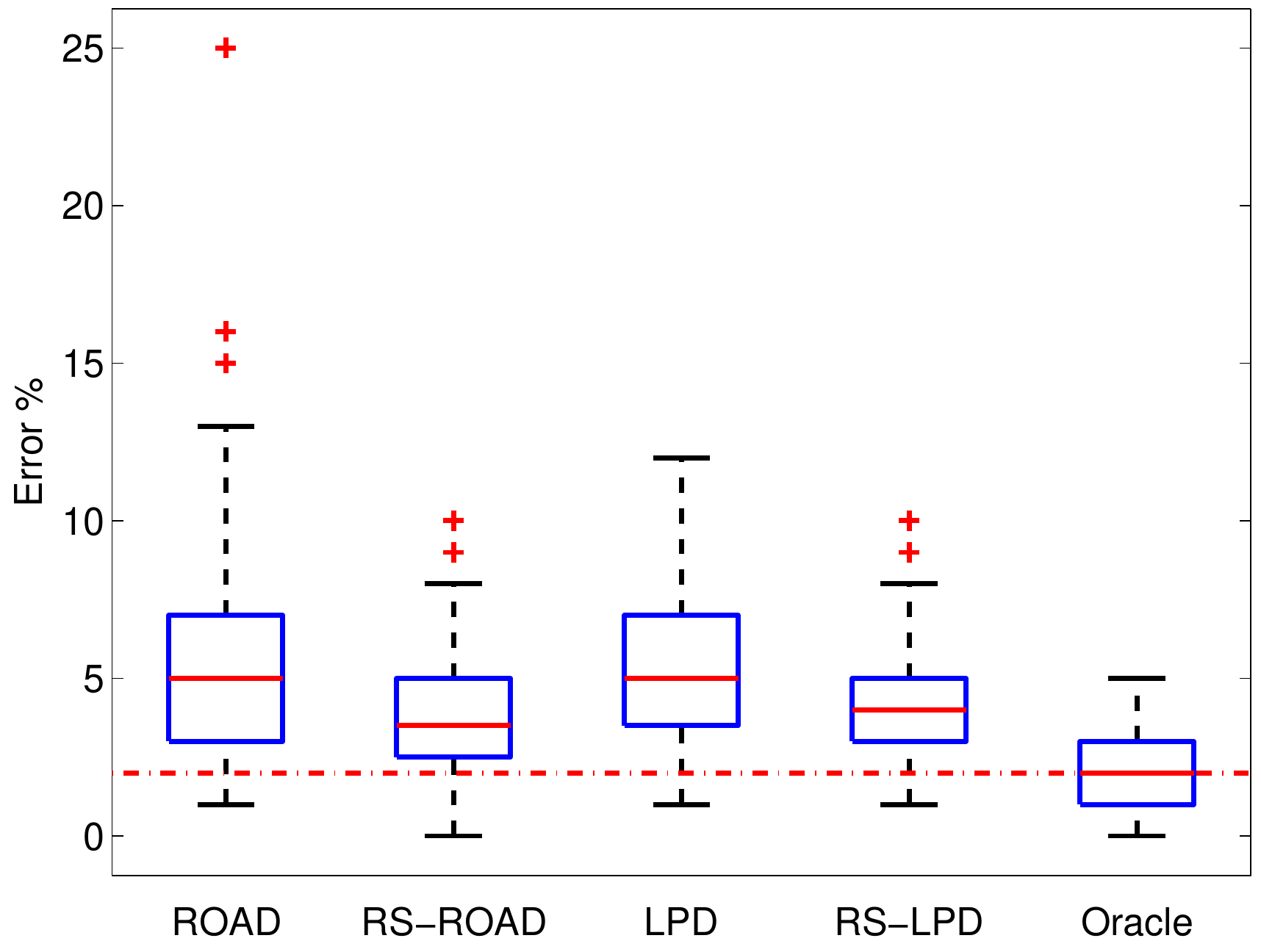}
    \includegraphics[width=2in]{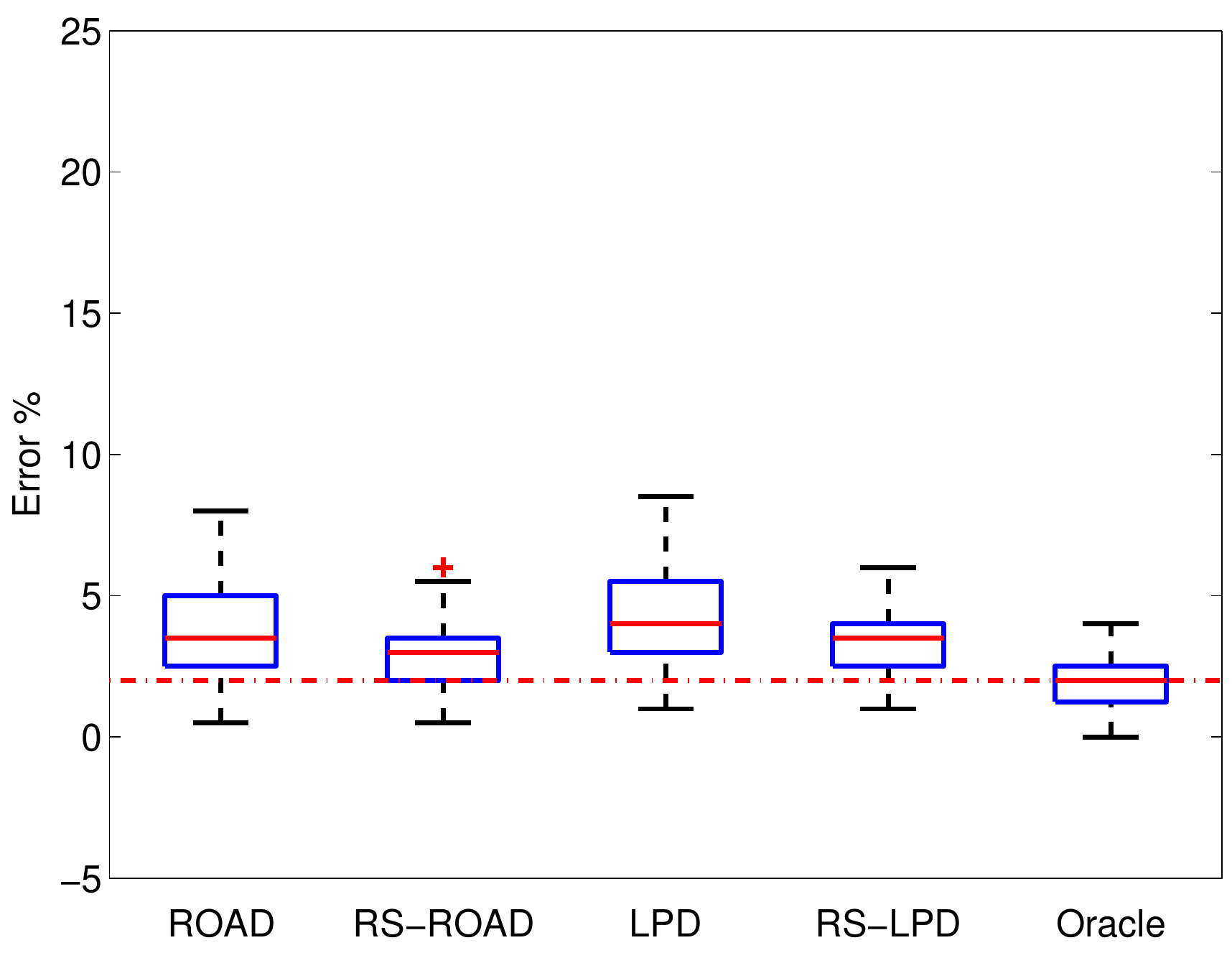}
    \includegraphics[width=2in]{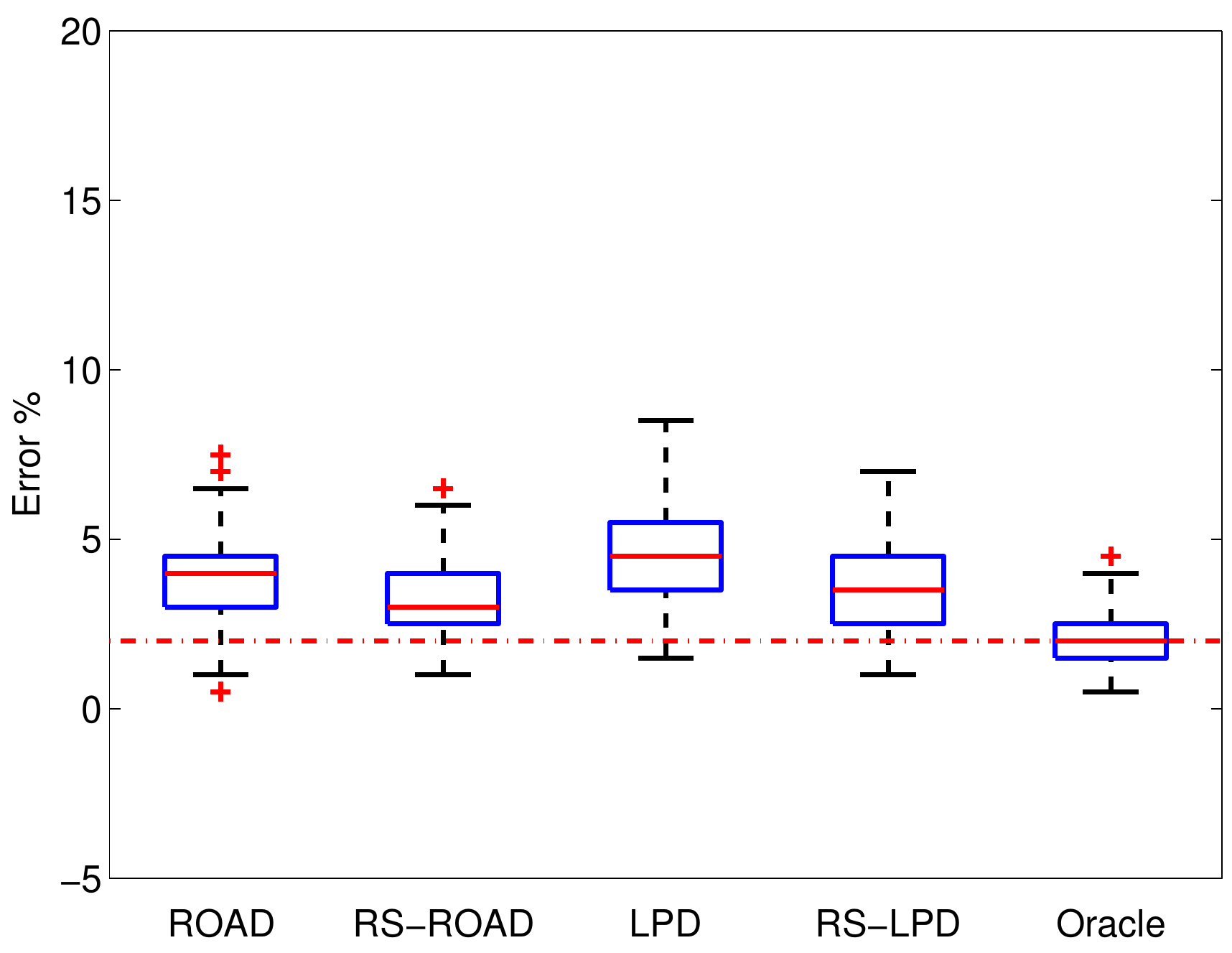}
\caption{Simulation results of Model 2: the boxplots from left to right correspond to the cases $(n_1,p)=(30,
200)$, $(50, 200)$, $(50, 400)$, $(100, 200)$ and $(100, 400)$.}\label{Fig:SecondSetOfModels:2}
\end{figure}

\begin{figure}[htp]
\centering
    \includegraphics[width=2in]{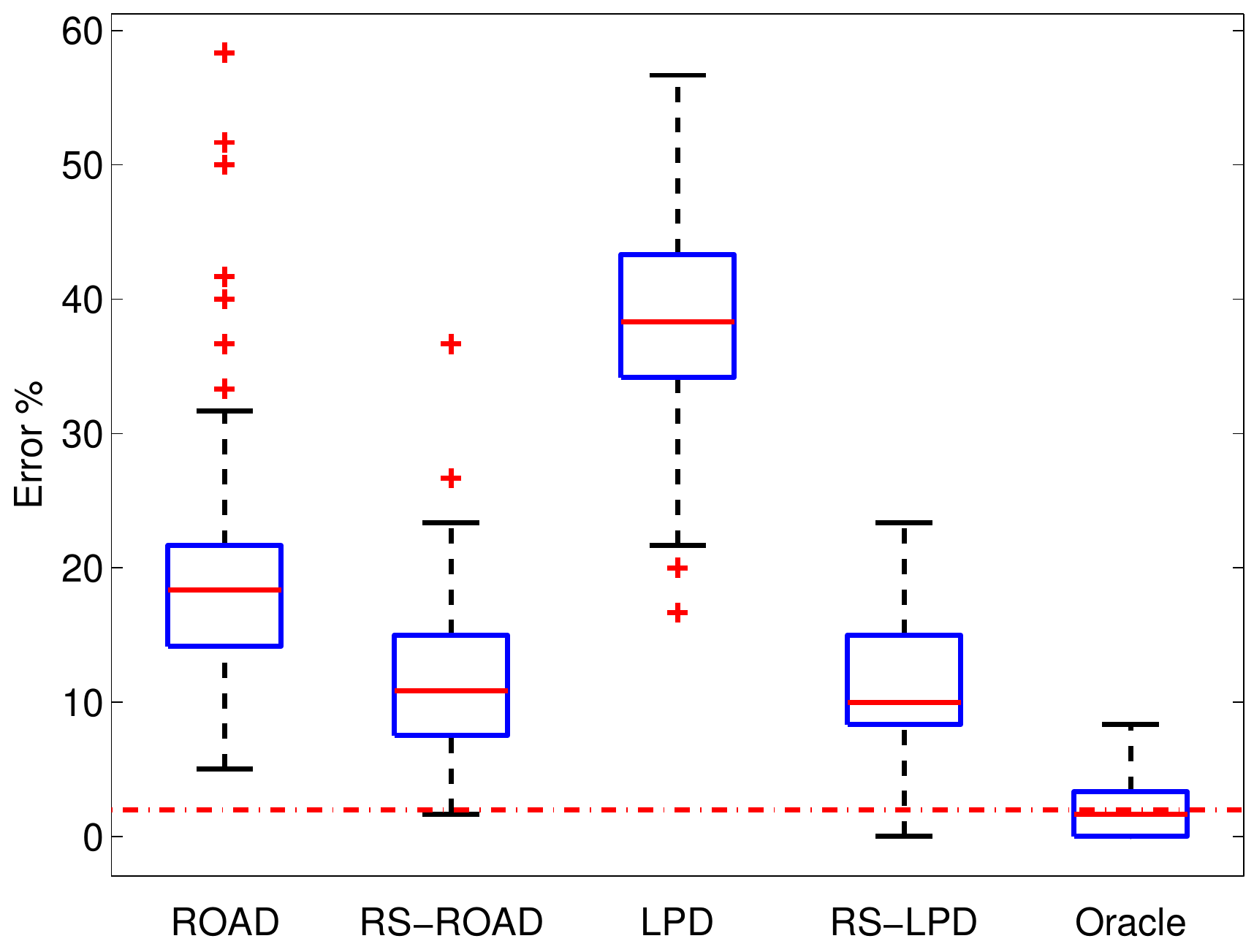}
    \includegraphics[width=2in]{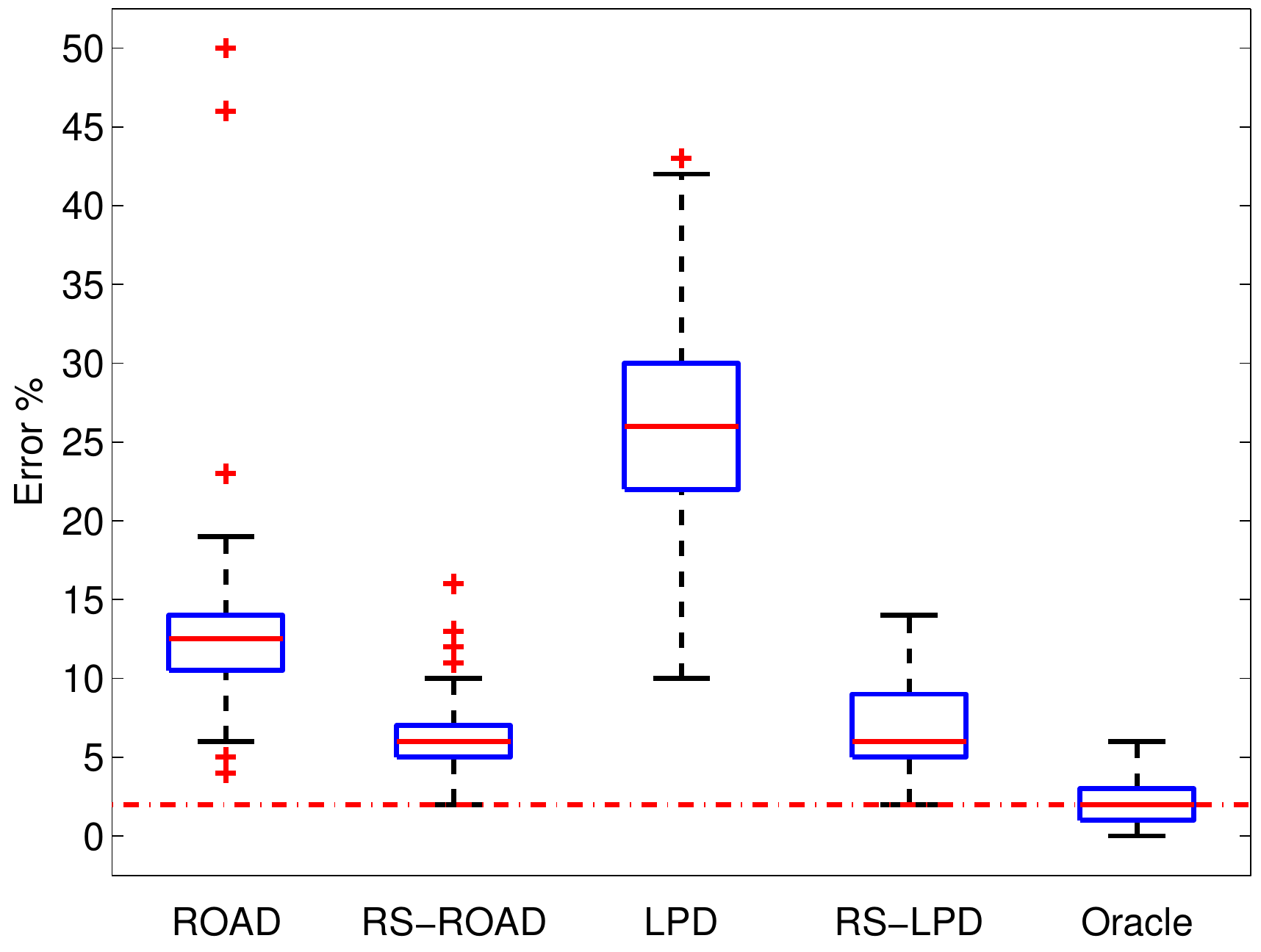}
    \includegraphics[width=2in]{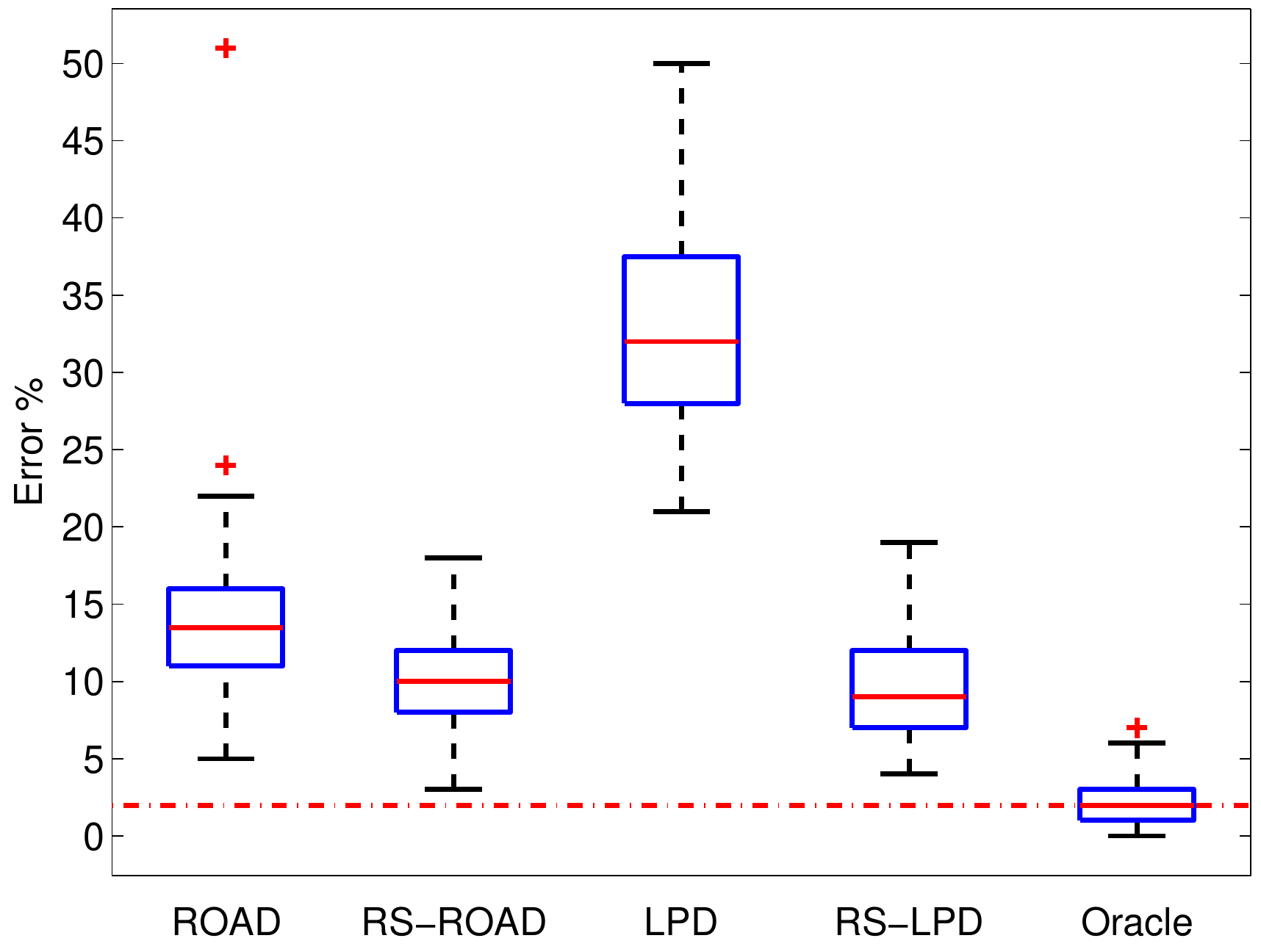}\\
    \includegraphics[width=2in]{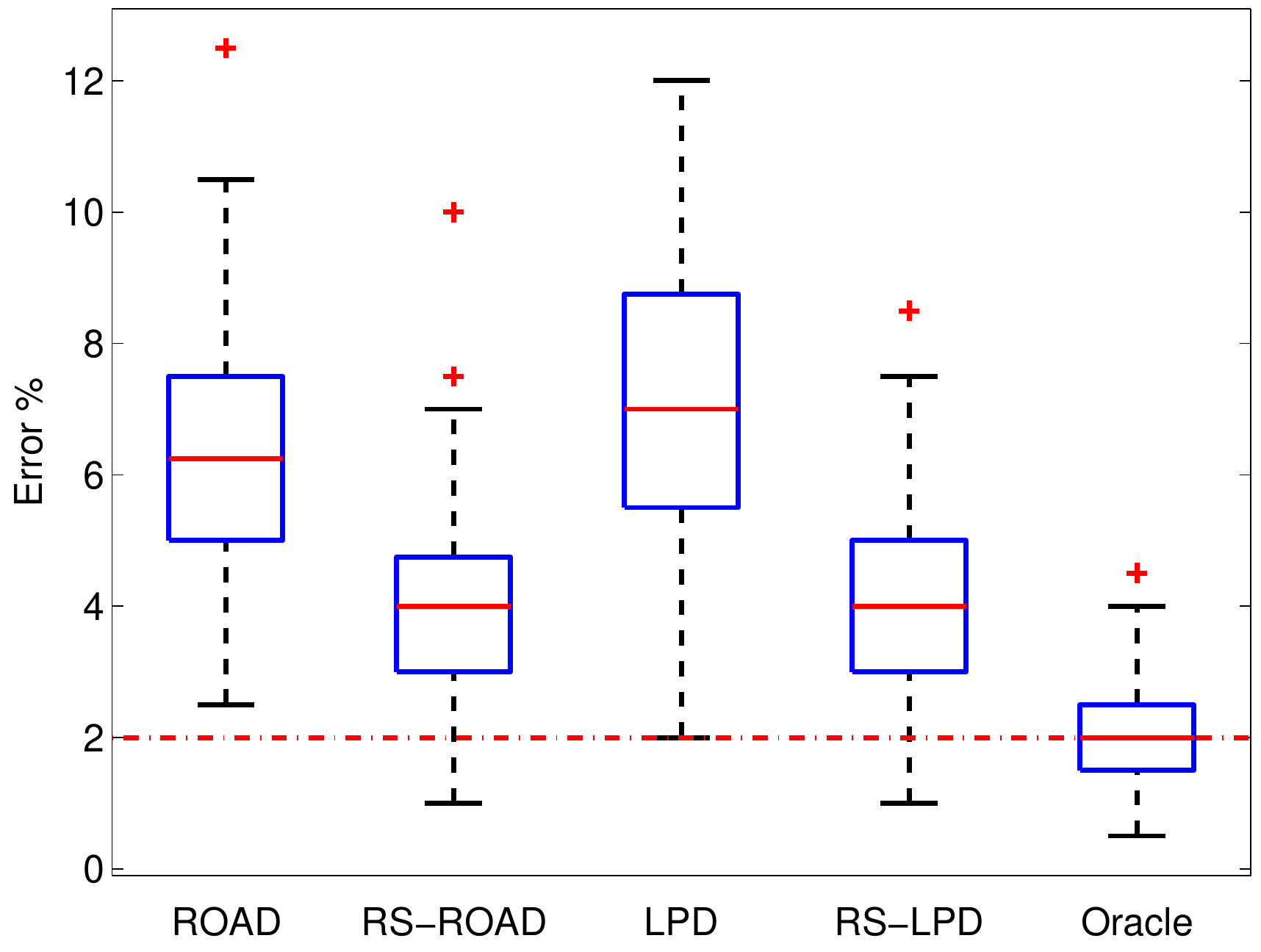}
    \includegraphics[width=2in]{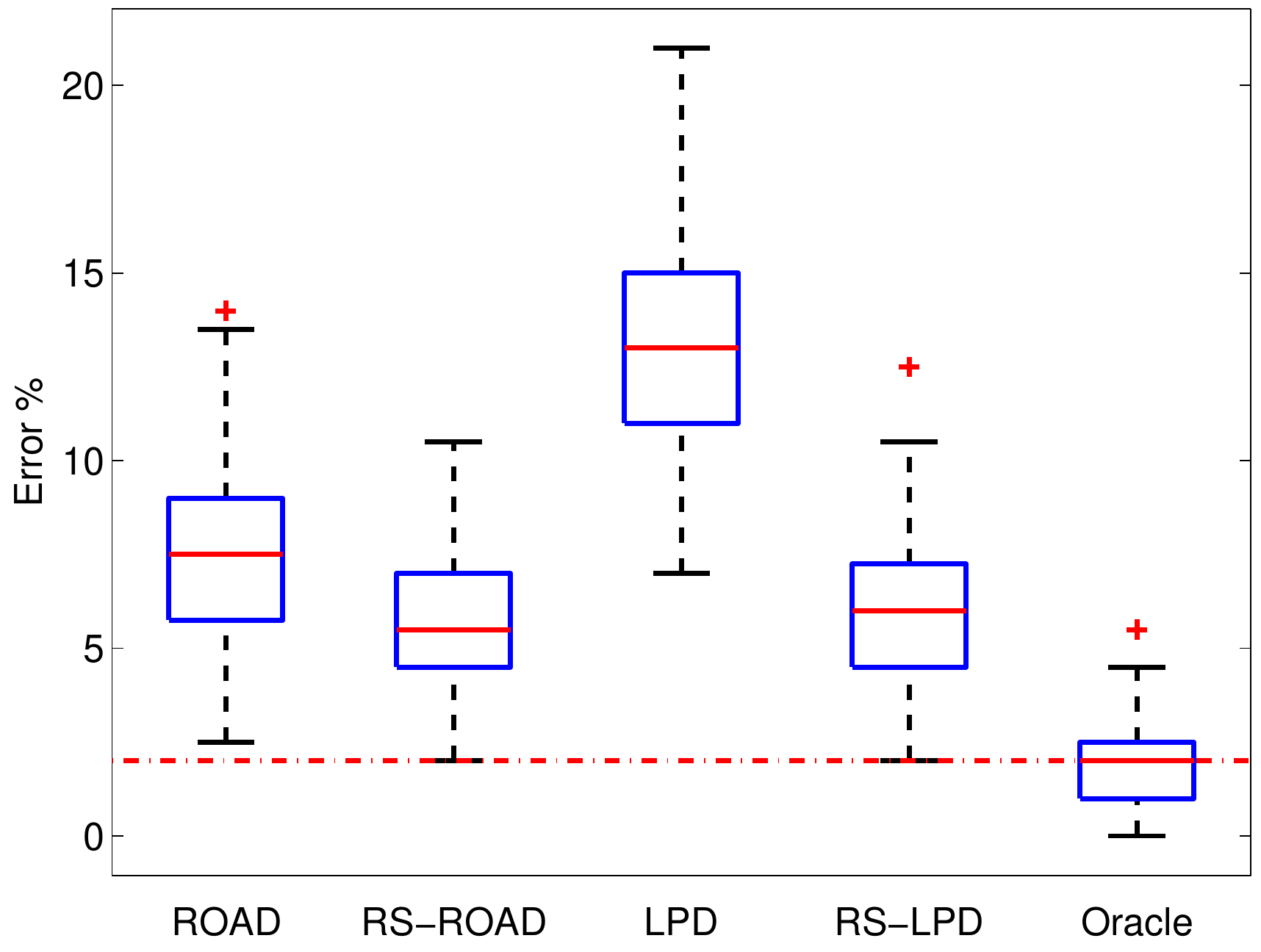}
\caption{Simulation results of Model 3: the boxplots from upper-left to lower-right correspond to the cases $(n_1,p)=(30,
200)$, $(50, 200)$, $(50, 400)$, $(100, 200)$ and $(100, 400)$.}\label{Fig:SecondSetOfModels:3}
\end{figure}

In our next numerical simulations, we consider the following three covariance structures:
\vspace{-0.2cm}

\begin{enumerate}
\item[] \textbf{Model 1:} $\bSigma=(\sigma_{i,j})$ with     $\sigma_{i,i}=1$ and $\sigma_{i,j}=0.5$ for $i\ne j$.

\item[] \textbf{Model 2:} $\bSigma=(\sigma_{i,j})$ with     $\sigma_{i,j}=0.7^{|i-j|}$.

\item[] \textbf{Model 3:} $\bSigma=\bI+\bA\bA^{\top}$ where $\bI$ is the identity matrix and $\bA$ is $p\times 5$ matrix with entries generated independently from $\cN(0,1)$.
\end{enumerate}
Without loss of generality, we set $\mu_1=\bm{0}$ and $\mu_2=(a\bm{1}^\top_{p/2},\ \bm{0}_{p/2}^\top)^\top$, where $a$ is chosen specifically for each model such that the expected classification error of the oracle rule is 2\%. Similar as before, for each simulation, we generate $2n_1$ independent observations for each class, where $n_1$ observations are used as training data and the other $n_1$ observations are used for testing. Results of Models 1-3 are presented in Figures \ref{Fig:SecondSetOfModels:1}-\ref{Fig:SecondSetOfModels:3} respectively, with various sample sizes and dimensionality. For Models 1 and 3 where the covariance structure is spiked, the improvement of the RS methods over the ROAD/LPD is remarkable. For Model 2 where the covariance structure is far from being spiked, the RS methods still generally outperform their counterparts.

In order to show that the improvement by applying RS is relatively general, 
we consider the following two scenarios with randomly generated covariance matrices
\begin{enumerate}
\item[] \textbf{Random Model 1:}
    $\bSigma=\left(\frac{M}{\|M\|}\right)^\top \left(\frac{M}{\|M\|}\right)+\mbox{diag}(v)$ with each entry of $p\times p$ matrix $M$
    being generated independently from $\mathcal{N}(0,1)$ and $v$ from $\mathcal{U}(0,1)$, where $\|M\|$ is the operator norm of $M$.

\item[] \textbf{Random Model 2:}
    $\bSigma=4\left(\frac{M}{\|M\|}\right)^\top \left(\frac{M}{\|M\|}\right)$ with each entry of $M$ being generated independently from $\mathcal{N}(0,1)$.
\end{enumerate}
We fix $n_1=30$ and $p=300$ and consider different sparsity levels of $\bbeta$ with ${||\bbeta||_0}/{p}=5\%, 10\%, \ldots, 95\%, 100\%$. We randomly generate $\bbeta$ with a given sparsity level , whose nonzero entries are IID from $\mathcal{N}(0,1)$. We then normalize $\bbeta$ such that $\bbeta^\top\bSigma\bbeta=12$. We fix $\bmu_1=\bm{0}$ and let $\bmu_2=-{\bSigma}\bbeta$. We repeat our data generation and classification 100 times for each scenario and record the average classification errors and their standard deviations.

\begin{figure}[htp]
\centering
    \includegraphics[width=2.5in]{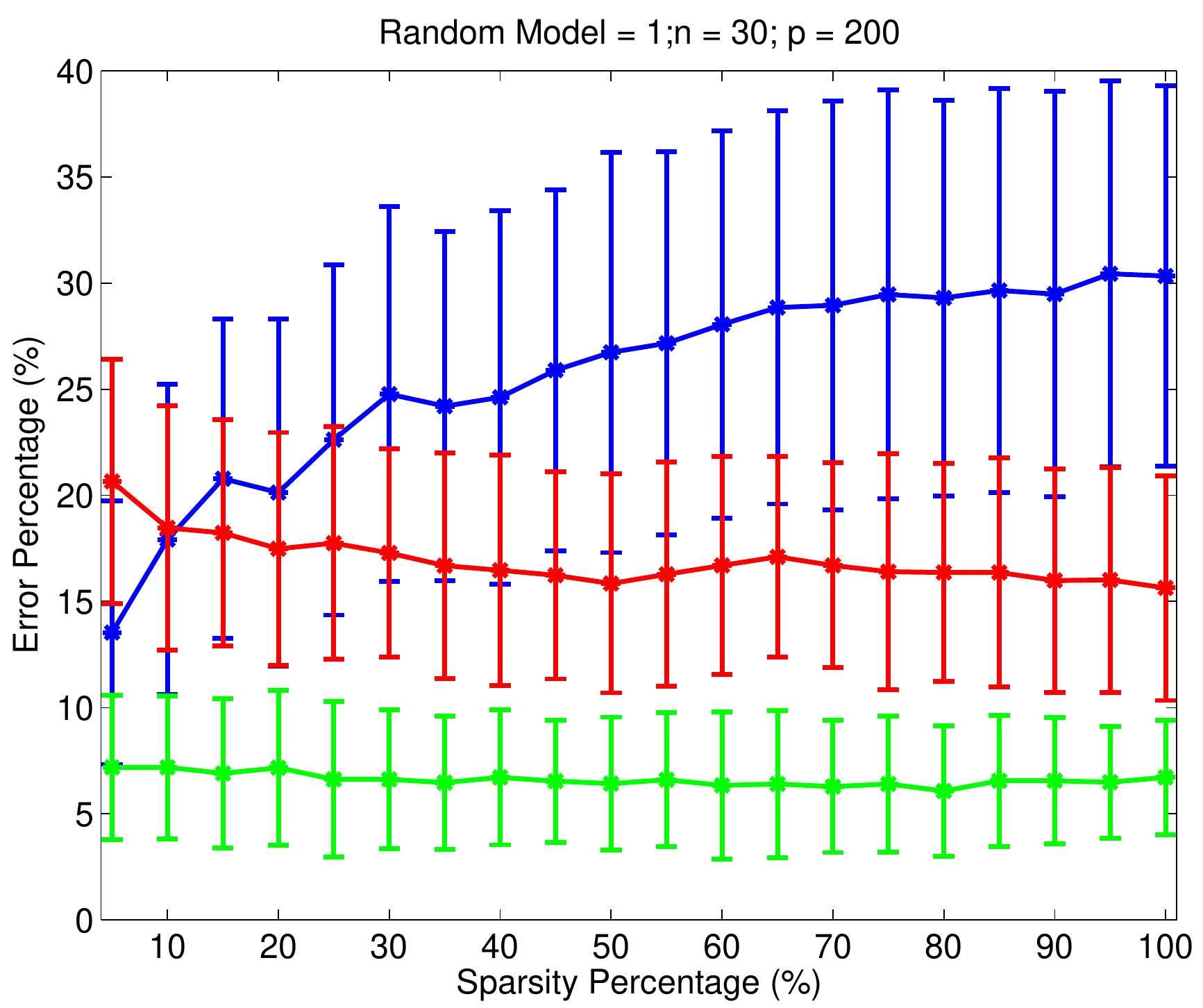}
    \includegraphics[width=2.5in]{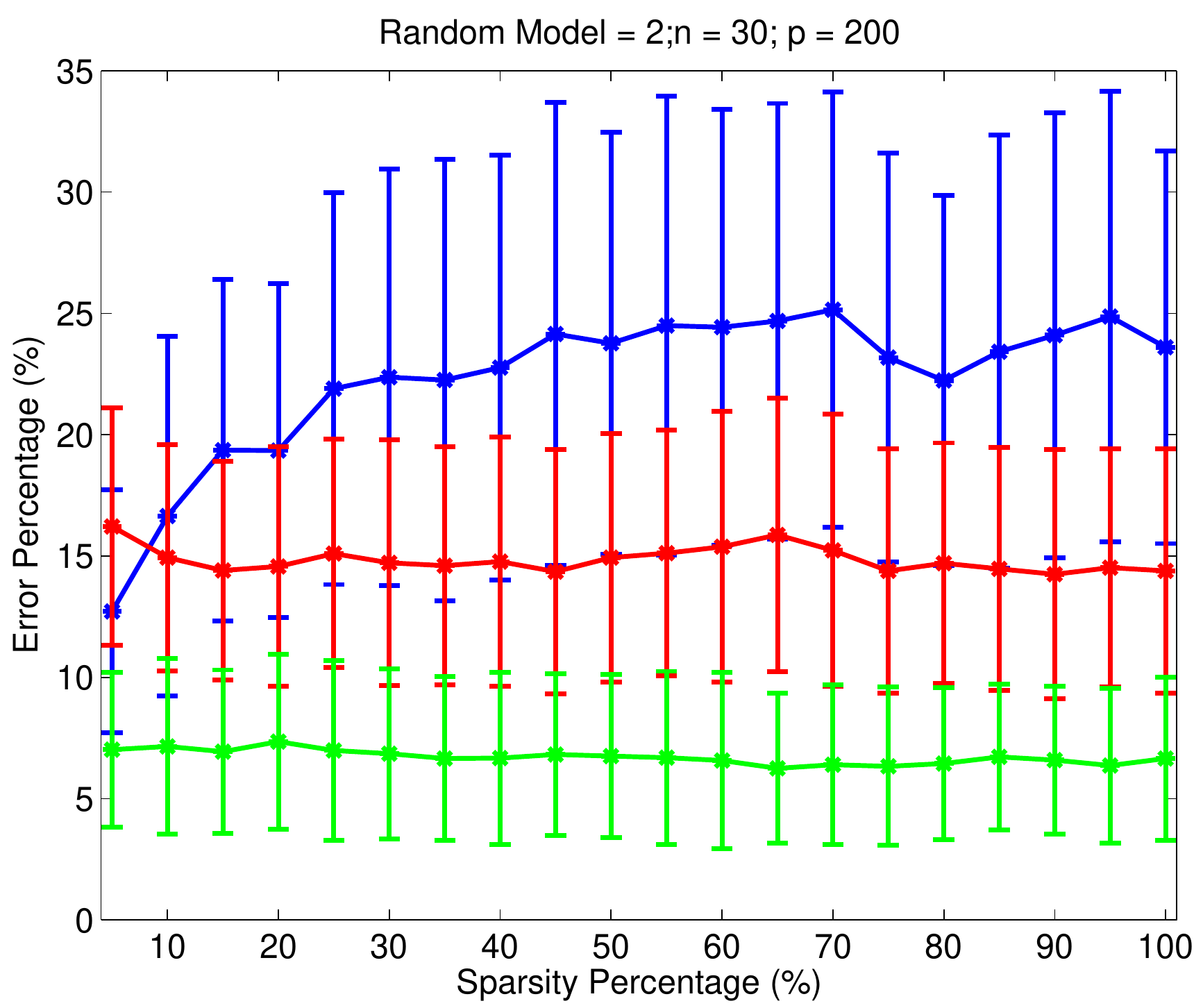}
\caption{Average classification errors of ``Random Model 1" (left) and ``Random Model 2" (right) for ${||\bbeta||_0}/{p}=5\%, 10\%, \ldots, 95\%, 100\%$, with
ROAD in blue, RS-ROAD in red and ORACLE in green.  Bars indicate the standard deviations of classification errors across 100 simulations.}\label{Fig:Sparsity:ROSE:ROAD:Random}
\end{figure}

We compare the results of ROAD and RS-ROAD which are shown in Figure \ref{Fig:Sparsity:ROSE:ROAD:Random}. As one can see when $\bbeta$ is very sparse, ROAD outperforms RS-ROAD as expected. However, the performance of ROAD highly depends on the sparsity level. On the other hand, RS-ROAD has significantly smaller overall error rates, and has the same qualitative behavior as the ORACLE. In particular, RS-ROAD is robust against the sparsity level of the true data generating procedure.

\subsection{Real Data: Leukemia and Lung Cancer}\label{Subsec:Cancer:Classification}

We now evaluate the performance of our proposed RS procedure on two popular
gene expression data set: Leukemia \citep{golub1999molecular} and lung
cancer \citep{gordon2002translation}. The two data sets come with separate
training and testing sets of data vectors. The Leukemia data set contains
$p=7129$ genes with $n_1=27$ acute lymphoblastic leukemia (ALL) and $n_2=11$
acute myeloid leukemia (AML) vectors in the training set. The testing set
includes 20 ALL and 14 AML vectors. The Lung Cancer data set contains
$p=12533$ genes with $n_1=16$ adenocarcinoma (ADCA) and $n_2=16$ mesothelioma
training vectors. The testing set has 134 ADCA and 15 mesothelioma vectors.

In our experiments, we put all the 47 (27 training $+$ 20 testing data) ALL
vectors and 25 (11 training $+$ 14 testing data) AML vectors together and
randomly select 23 ALL and 12 AML as training and the rest as testing. We
repeat the experiments 20 times. We conduct a similar experiment on Lung
cancer data by randomly select 75 ADCA and 15 mesothelioma data vector as
training and the rest as testing, and repeat 20 times. The classification
results of the aforementioned experiments using IR, NSC, ROAD and RS-ROAD are
presented in Table \ref{Table:Cancer}, where RS-ROAD has the best overall performance.

\begin{table}[htp]
\caption{{Classification errors for cancer data.}}\label{Table:Cancer}
\centering
\begin{tabular}{c|c|c|c|c}
\hline
\multicolumn{1}{c|}{Errors \% (std \%)} & \multicolumn{1}{c|}{IR} & \multicolumn{1}{c|}{NSC} &
\multicolumn{1}{c|}{ROAD} & \multicolumn{1}{c}{RS-ROAD}\\
\hline
 \multicolumn{1}{c|}{Leukemia}         & 4.2708 (2.9998) &  8.5135 (8.4232) & 6.3514 (5.9650) & 4.4595 (3.0721) \\
 \multicolumn{1}{c|}{Lung Cancer}      & 3.4669 (1.4381) & 10.4396 (7.2675) & 1.3736 (1.0621) & 0.9341 (0.8931) \\
 \hline
\end{tabular}
\end{table}

\subsection{Real Data: Shape Classifications}\label{Subsec:Shape:Classification}

\begin{figure}[htp]
\centering
    \includegraphics[width=3.5in]{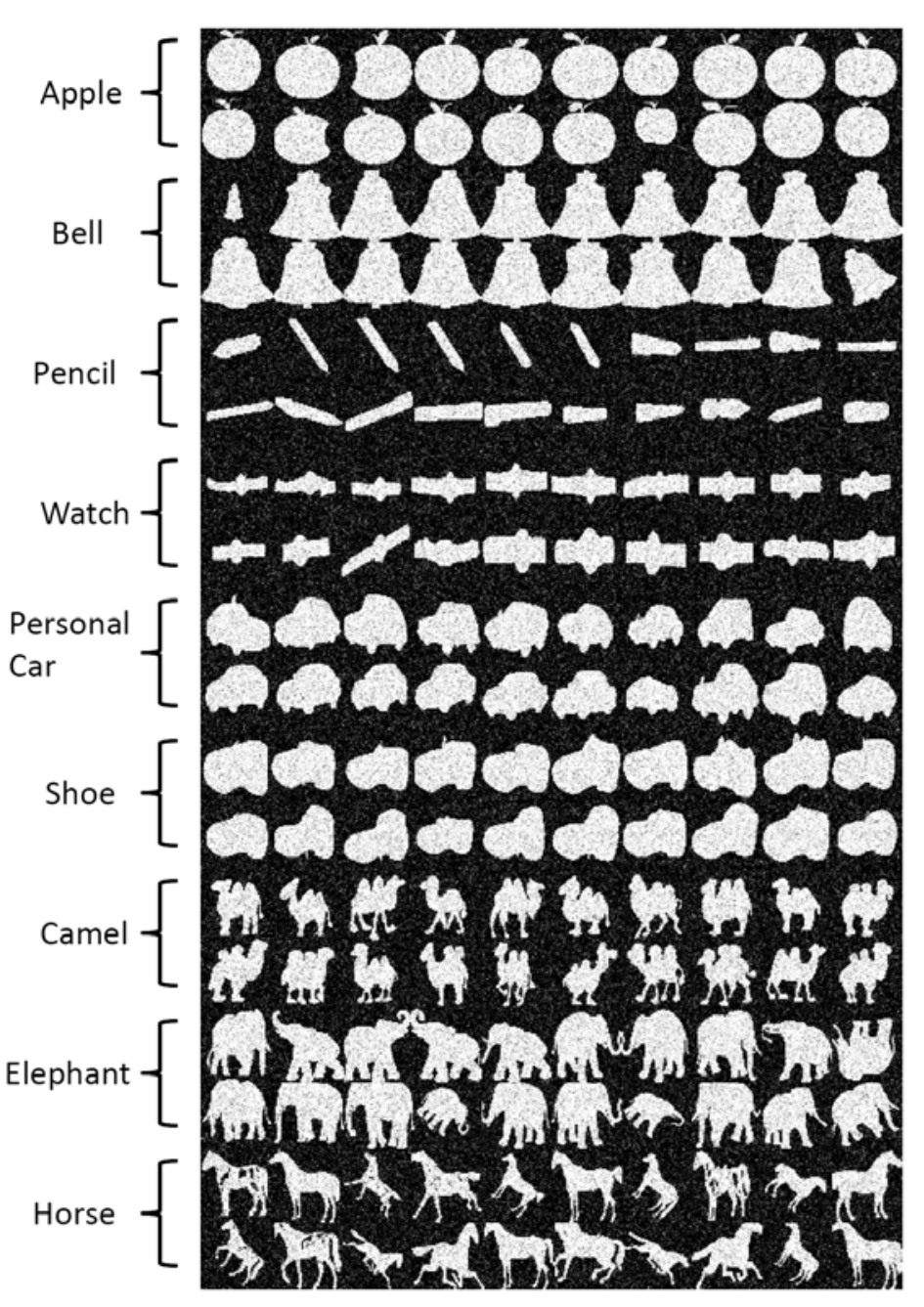}
\caption{Selected shape images: resized to $50\times50$ with additive Gaussian noise.}\label{Fig:Shape:Images}
\end{figure}

We also evaluate the performance of RS on shape classification, which is
one of the most fundamental and important problems in computer vision and
machine learning. All the shapes are represented by 2D binary images. We
downloaded the MPEG-7 CE Shape-1 Part-B data set \citep{thakoor2007hidden}
and selected a subset of it for our tests. Since the images in the dataset
generally have different sizes, we resized them to the same size $50\times50$
(i.e., $p=2500$) using the Matlab command \texttt{imresize} with bi-cubic
interpolation. All the selected and resized shape images are shown in Figure
\ref{Fig:Shape:Images}.

There are 20 images for each shape class. After being loaded, each image is a matrix, with elements taking integer values in $[0, 255]$. In order to test the robustness of
the classifiers, we also added Gaussian noise $N(0,50^2)$ to all the selected images. 
For every pair of shapes, we randomly select 10 from each class as testing data and the rest as training data (i.e., $n_1=n_2=10$). We repeat this 50 times for each of the shape pairs. The average classification errors by IR, NSC, ROAD and RS-ROAD are summarized in Table \ref{Table:Shapes}.  We observe that RS-ROAD has the best overall performance, and it consistently improves ROAD in all scenarios.

\begin{table}[htp]
\caption{Classification errors for shapes.}\label{Table:Shapes} \centering
\begin{tabular}{c||c|c|c|c|c|c|c|c}
\hline
\multicolumn{1}{c||}{Errors (\%)} & \multicolumn{2}{c|}{IR} & \multicolumn{2}{c|}{NSC} &
\multicolumn{2}{c|}{ROAD} & \multicolumn{2}{c}{RS-ROAD}
\\\cline{2-9}
 Shape Pairs:
 & \multicolumn{1}{c|}{mean} & \multicolumn{1}{c|}{std} & \multicolumn{1}{c|}{mean} & \multicolumn{1}{c|}{std} & \multicolumn{1}{c|}{mean} & \multicolumn{1}{c|}{std}& \multicolumn{1}{c|}{mean} & \multicolumn{1}{c}{std}\\
\hline
 \multicolumn{1}{c||}{Apple \& Bell}            & 7.9  & 3.0 &  7.7 & 3.1 &  8.3 &  4.5 &  7.8 & 3.4  \\
 \multicolumn{1}{c||}{Pencil \& Watch}          & 19.2 & 6.1 & 20.4 & 7.1 & 18.2 &  6.9 & 16.0 & 7.1  \\
 \multicolumn{1}{c||}{Personal Car \& Shoe}     & 7.7  & 5.1 & 11.3 & 6.6 & 13.2 &  6.9 &  6.1 & 4.2  \\
 \multicolumn{1}{c||}{Camel \& Elephant}        & 8.8  & 6.6 & 12.1 & 9.1 & 20.3 & 11.0 &  6.9 & 4.0  \\
 \multicolumn{1}{c||}{Camel \& Horse}           & 9.6  & 7.1 & 11.8 & 9.5 & 22.0 & 10.6 &  7.7 & 5.6  \\
 \multicolumn{1}{c||}{Elephant \& Horse}        & 8.8  & 6.4 & 11.9 & 8.4 & 15.1 & 10.1 &  6.9 & 5.9  \\
 \hline
\end{tabular}
\end{table}

\subsection{Choice of $\rho$}

Here we shall mainly discuss two issues related to the choice of $\rho$ in $\bm\Sigma^{tot}_{\rho}$: (1) the sensitivity of the classification results to the choices of $\rho$; (2) data-adaptive selection of $\rho$ by cross-validation.

\subsubsection{Sensitivity to $\rho$}

\begin{figure}[htp]
\centering
    \includegraphics[width=2.0in]{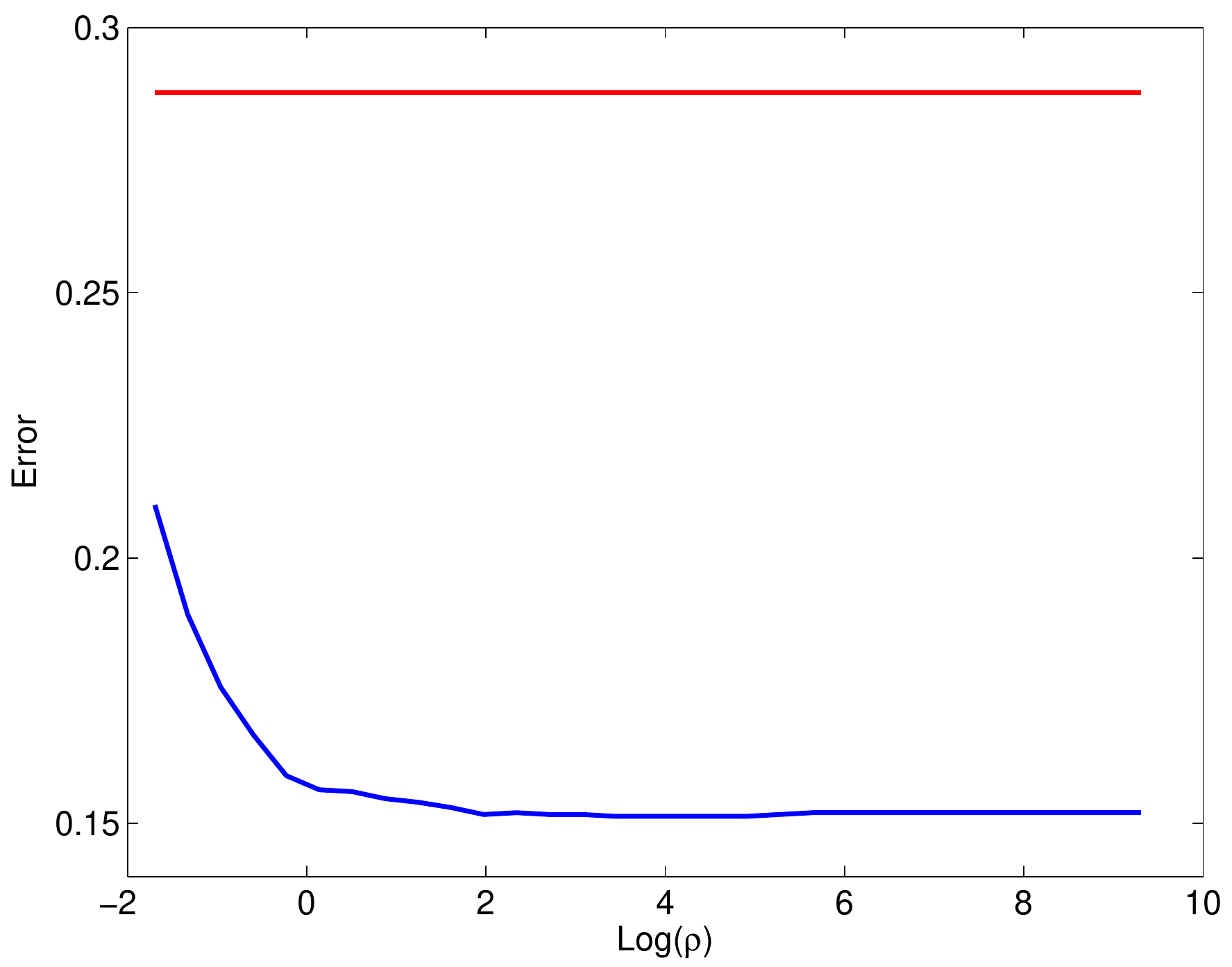}
    \includegraphics[width=2.0in]{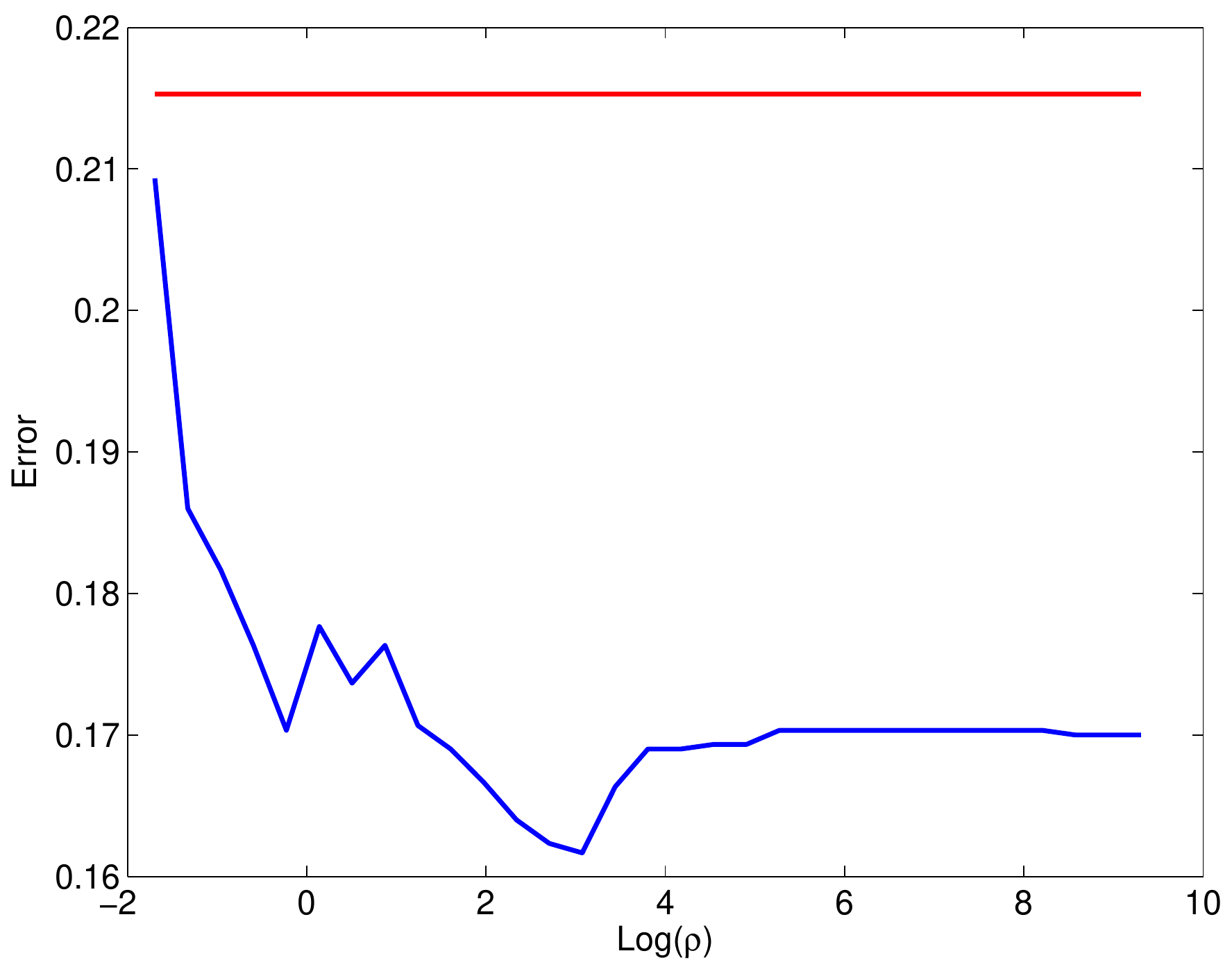}
    \includegraphics[width=2.0in]{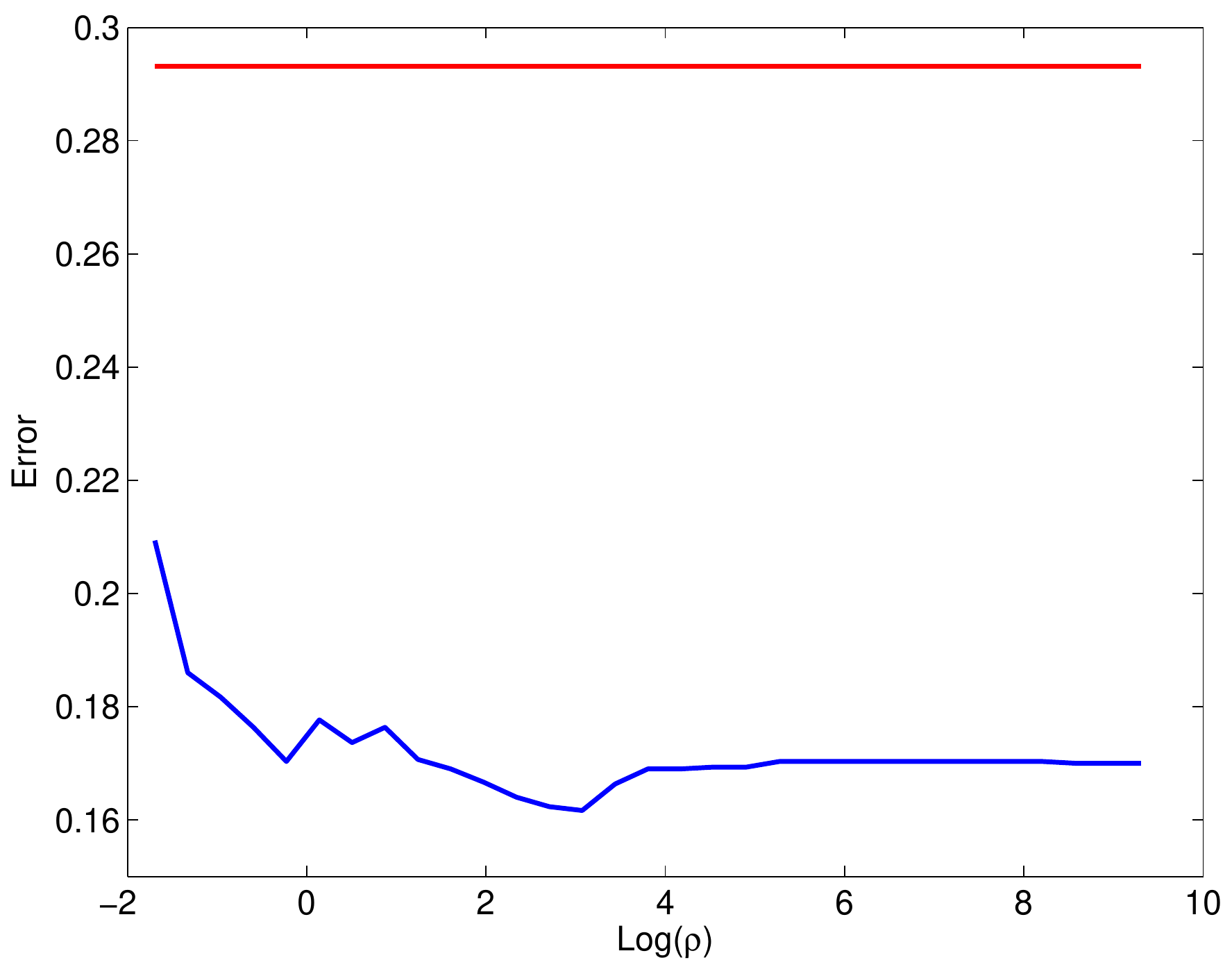}\\
    \includegraphics[width=2.0in]{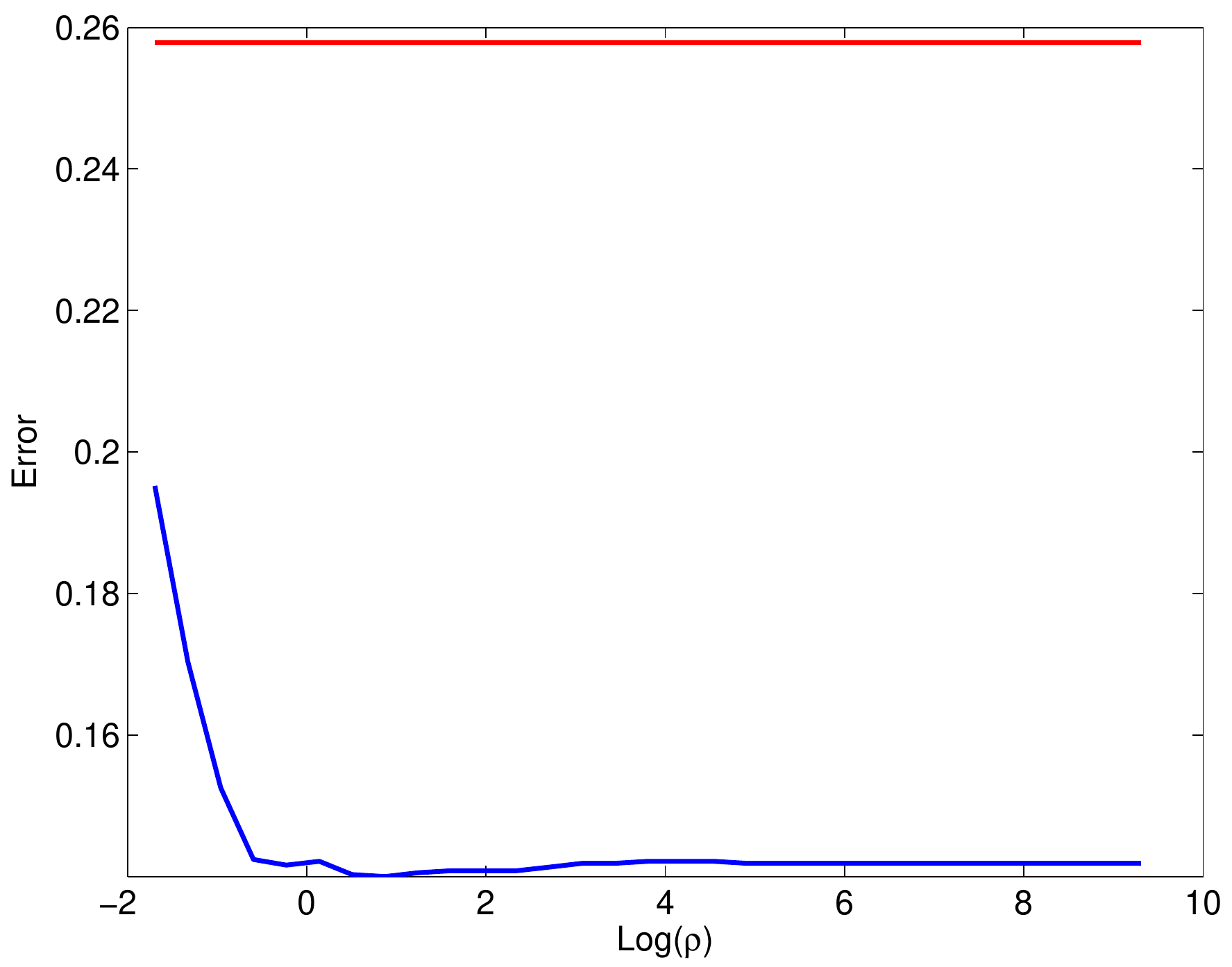}
    \includegraphics[width=2.0in]{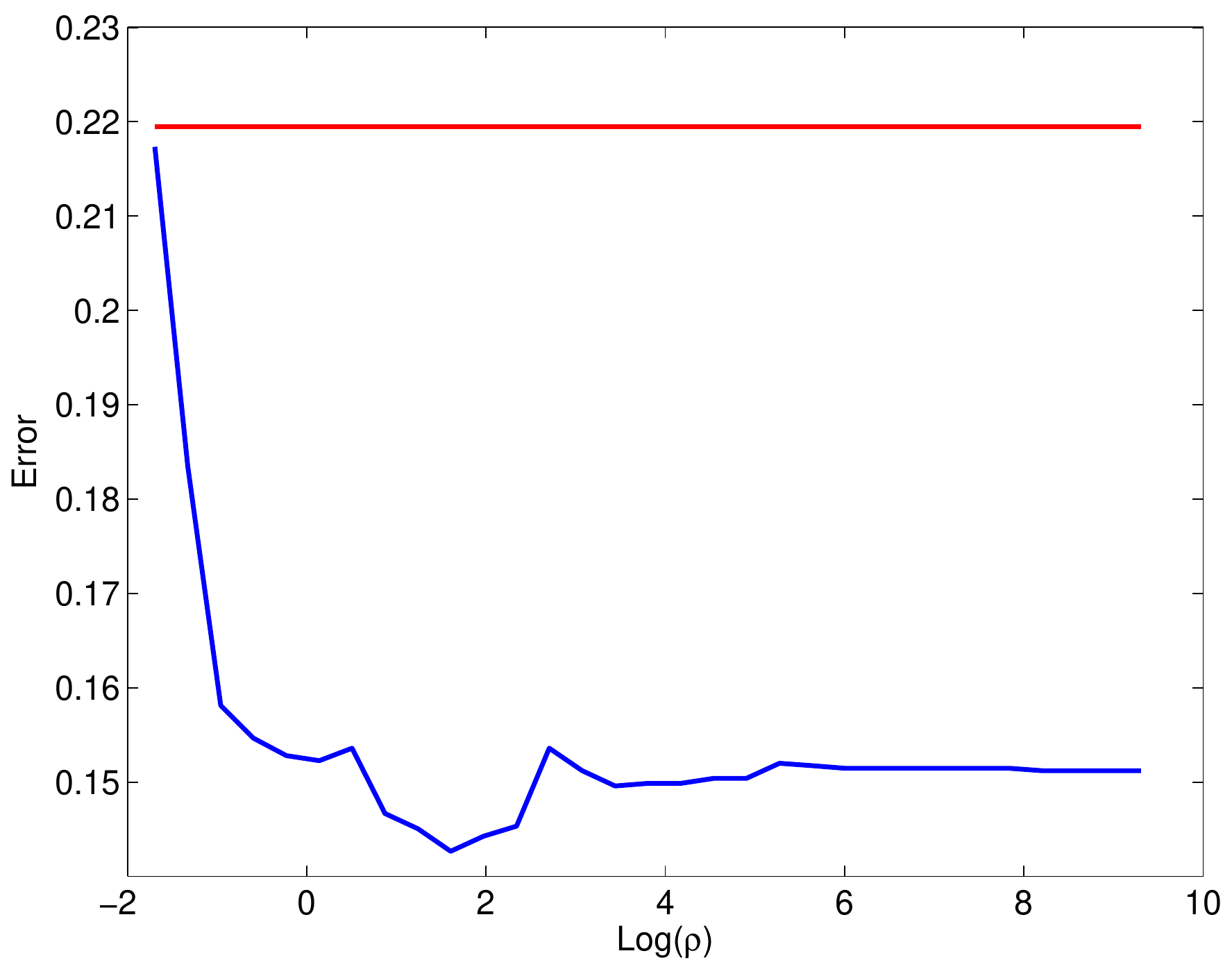}
    \includegraphics[width=2.0in]{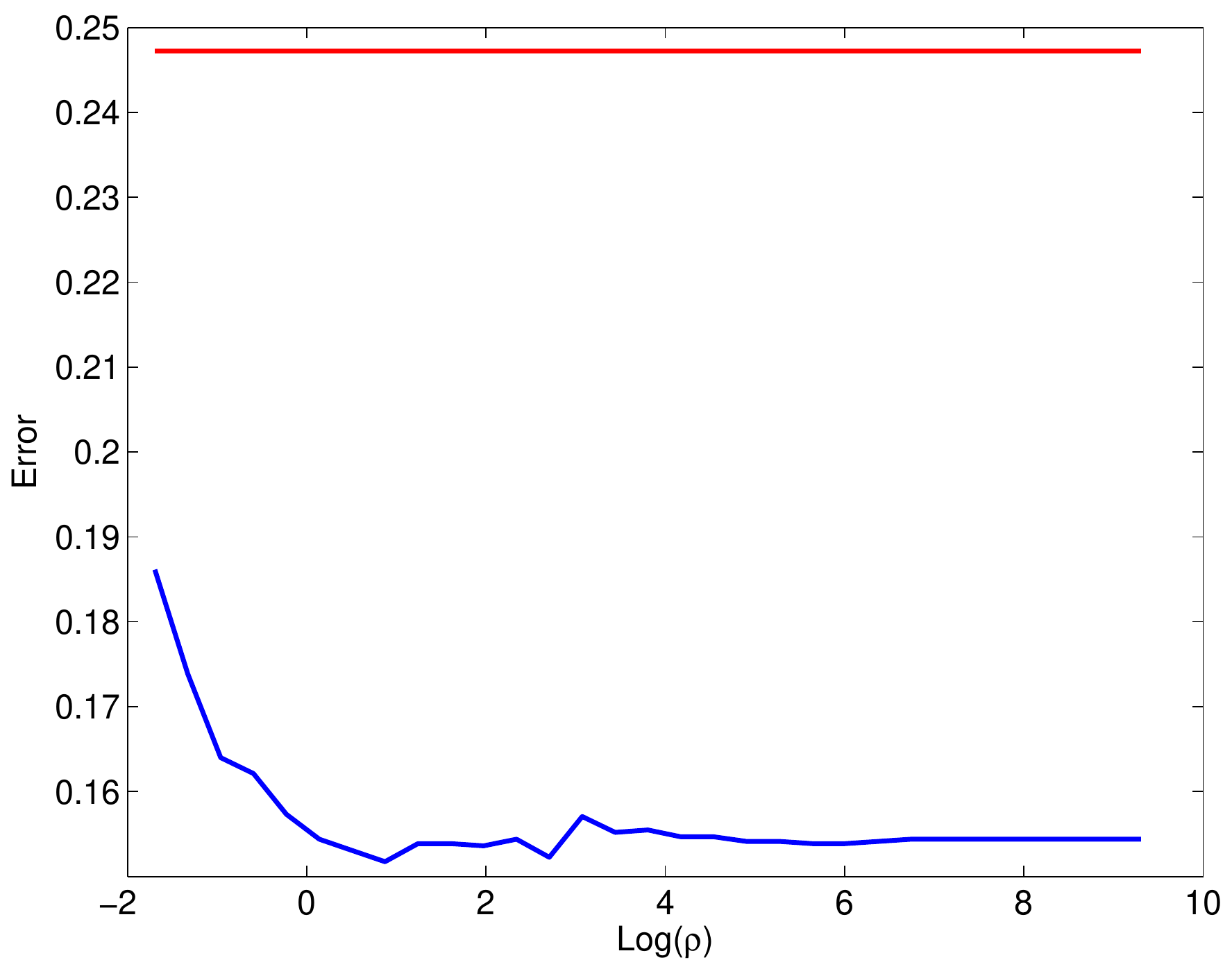}
\caption{Classification errors of RS-ROAD with various $\rho$ (blue curves) v.s. ROAD (red lines). Plots in the first row correspond to the case $n_1=n_2=30$ and plots in the second row correspond to $n_1=30$ and $n_2=45$. Columns 1-3 correspond to the Toy Models 1-3.}\label{Fig:Varied:Rho}
\end{figure}

In the following simulations, we take the toy models 1-3 with $a_i$'s chosen such that the oracle error rate is $10$\%, and use the method RS-ROAD as an example. Let $\hat\bU_{\rho}$ be the eigenvectors of $\hat\bSigma+\rho\hat\bdelta\hat\bdelta^\top$ with various values of $\rho$. The average classification errors (among 100 replicates) of RS-ROAD with various $\rho$ are shown in Figure \ref{Fig:Varied:Rho}, where the blue curves show the errors associated to $\rho$ and the red horizontal lines indicate the errors of ROAD. As we can see, the best choice of $\rho$ depends on the scenario. Although it seems that choosing $\rho$ optimally is a complicated issue, the plots in Figure \ref{Fig:Varied:Rho} do indicate that for a large range of $\rho$, the classification results have significant improvements over a non-rotated classifier such as ROAD. This also indicates the robustness of the RS procedure to the choices of the parameter $\rho$. In general, any reasonable positive value of $\rho$  should work well in most applications (Figure 9 shows the workable range of $\log \rho \in [-1,10]$), if one does not have the resources or time to perform cross-validation.

\subsubsection{Cross-Validation choice of $\rho$}

Cross-validation on $\rho$ is computationally expensive when $p$ is large. See Remark 4 for reduction of computation.  When $\bm\Sigma$ has a (quasi)-spiked covariance structure, i.e. there are $k$ eigenvalues that are significantly larger than the rest $p-k$ eigenvalues, and if $k$ is much less than the number of observations $n$, then we may use $\tilde\bU$ to rotate the data instead of using $\hat\bU$. Recall that $\tilde\bU$ is the collections of the $n$ eigenvectors of $\hat{\bm\Sigma}^{tot}$ corresponding to the $n$ largest eigenvalues. Then after rotating the data using $\tilde\bU$, we reduce the dimension of the problem from $p$ to $n$ which will be significant reduction when $n\ll p$ (e.g. the real data considered in the previous two sections).  We can also take $\tilde\bU$ to be principal components, with dimensionality much less than $n$.

Our first simulations show that using $\tilde\bU$ instead of $\hat\bU$ does not hurt the classification error. We take the toy model 1-3 with $a_i$'s chosen such that the oracle error rate is $10$\%, and use the method RS-ROAD as an example. We set $n_1=n_2=10$ (i.e. $n=20$) and $p=50$. The results are summarized in Table \ref{Table:Reduced:Eigenvectors}.

\begin{table}[htp]
\caption{Classification errors and their standard deviations.}\label{Table:Reduced:Eigenvectors}
\centering
\begin{tabular}{c||c|c|c}
\hline
\multicolumn{1}{c||}{Errors \% (std \%)} &
\multicolumn{1}{c|}{Toy Model 1} & \multicolumn{1}{c|}{Toy Model 2}& \multicolumn{1}{c}{Toy Model 3}\\
\hline\hline
 \multicolumn{1}{c||}{Using $\hat\bU$}           & 24.9500 (11.3817) & 26.8500 (12.6861) & 26.7000 (12.5171)\\
 \multicolumn{1}{c||}{Using $\tilde\bU$}         & 25.0000 (11.5470) & 26.5000 (12.5831) & 26.9500 (12.5508)\\
 \hline
\end{tabular}
\end{table}

The previous simulation shows that we can reduce the size of the problem from $p$ to $n$ without sacrificing much of the classification quality. Since the computation cost can be greatly reduced in this way, cross-validation on $\rho$ is now a computationally viable approach. In our next experiments, we take the data of Leukemia and Lung cancer in Section \ref{Subsec:Cancer:Classification}, and conduct a similar experiment as we did before, except that we use $\tilde\bU$ and choose $\rho$ using 5-folds cross-validation. The classification results are summarized in Table \ref{Table:Cancer:Reduced}, where we also reproduce the results in Table \ref{Table:Cancer} for comparison. We also presented therein the average values of $\rho$ chosen by cross-validation along with their standard deviations. We repeat the same simulation to the shape data we presented in Section \ref{Subsec:Shape:Classification} and present comparisons and the estimated values of $\rho$ in Table \ref{Table:Shapes:Reduced}. As one can see that the choice of $\rho$ is generally different for different type of data, and using cross-validation to select $\rho$, we can further reduce the classification errors.

\begin{table}[htp]
\caption{Classification errors and their standard deviations for Leukemia and Lung cancer.}\label{Table:Cancer:Reduced}
\centering
\begin{tabular}{c||c|c}
\hline
\multicolumn{1}{c||}{Errors \% (std \%)} &
\multicolumn{1}{c|}{Leukemia Cancer} & \multicolumn{1}{c}{Lung cancer}\\
\hline
 \multicolumn{1}{c||}{W/O Cross-Validation}           & 4.4595 (3.0721) & 0.9341 (0.8931) \\
 \multicolumn{1}{c||}{Cross-Validation}               & 4.0541 (3.9702) & 0.6593 (0.7479) \\
 \multicolumn{1}{c||}{Estimated $\rho$}           & 0.2241 (0.2630) & 0.0848 (0.0890) \\ \hline
\end{tabular}
\end{table}

\begin{table}[htp]
\caption{Classification errors for shapes and the average values of $\rho$.}\label{Table:Shapes:Reduced} \centering
\begin{tabular}{c||c|c|c}
\hline
\multicolumn{1}{c||}{Shape Pairs} &
\multicolumn{1}{c|}{W/O Cross-Validation} & \multicolumn{1}{c|}{Cross-Validation} & \multicolumn{1}{c}{Estimated $\rho$}
\\\cline{2-4}
 &
\multicolumn{1}{c|}{error mean (std)} & \multicolumn{1}{c|}{error mean(std)} & \multicolumn{1}{c}{mean(std)} \\
\hline
 \multicolumn{1}{c||}{Apple \& Bell}                &  7.8 (3.4) &  7.4 (3.4) & 0.0806 (0.0810) \\
 \multicolumn{1}{c||}{Pencil \& Watch}              & 16.0 (7.1) & 15.0 (6.5) & 0.2561 (0.2757) \\
 \multicolumn{1}{c||}{Personal Car \& Shoe}         &  6.1 (4.2) &  5.5 (3.4) & 0.0639 (0.0808) \\
 \multicolumn{1}{c||}{Camel \& Elephant}            &  6.9 (4.0) &  7.1 (4.1) & 0.0723 (0.0763) \\
 \multicolumn{1}{c||}{Camel \& Horse}               &  7.7 (5.6) &  6.0 (6.1) & 0.1196 (0.1829) \\
 \multicolumn{1}{c||}{Elephant \& Horse}            &  6.9 (5.9) &  6.5 (4.9) & 0.0667 (0.0725) \\
 \hline
\end{tabular}
\end{table}

\noindent \textbf{Acknowledgements}

The authors are grateful to the Joint Editor, the Associate Editor and two referees for helpful comments, which have led to many improvements. Hao's research is supported by National Science Foundation grant DMS-1309507 and AMS-Simons Travel Grant. Fan's research is supported by the National Institute of General Medical Sciences of the National Institutes of Health through Grant Numbers R01-GM072611-9 and R01-GMR01GM100474 and National Science Foundation grants DMS-1206464 and DMS-1406266.

\section{Appendix}

{\bf Proof of Theorem 1}:  Let $a=\lambda_p>0$ and $a_i=\lambda_i-\lambda_p>0$. 
It then follows directly from Condition 1 and the singular value decomposition that
\begin{eqnarray}\label{a1}
\bSigma&=& a\bI+\sum_{i=1}^ka_i\bxi_i\bxi_i^{\top}
\end{eqnarray}
and
\begin{eqnarray}\label{a2}
\bSigma^{tot}&=& a\bI+\rho\bdelta\bdelta^{\top}+\sum_{i=1}^ka_i\bxi_i\bxi_i^{\top}.
\end{eqnarray}

It can be shown that
\begin{eqnarray}\label{a0}
(a\bI+\sum_{i=1}^ka_i\bxi_i\bxi_i^{\top})^{-1}=a^{-1}\bI-\sum_{i=1}^k\frac{a_i}{a(a+a_i)}\bxi_i\bxi_i^{\top}.
\end{eqnarray}
This can be directly verified by
\begin{eqnarray*}
& &(a\bI+\sum_{i=1}^ka_i\bxi_i\bxi_i^{\top})(a^{-1}\bI-\sum_{i=1}^k\frac{a_i}{a(a+a_i)}\bxi_i\bxi_i^{\top})\\
&=&\bI+\sum_{i=1}^ka^{-1}a_i\bxi_i\bxi_i^{\top}-\sum_{i=1}^k\frac{a_i}{a+a_i}\bxi_i\bxi_i^{\top}-\sum_{i=1}^k
\frac{a^2_i}{a(a+a_i)}\bxi_i\bxi_i^{\top}\\
&=&\bI,
\end{eqnarray*}
using the orthogonality
\[
    \bxi_i^{\top}\bxi_j =\left\{
                                           \begin{array}{ll}
                                             0, & i\ne j; \\
                                             1, & i=j.
                                           \end{array}
                                         \right.
\]


By (\ref{a0}),
\begin{eqnarray}
\bbeta=\bSigma^{-1}\bdelta=a^{-1}\bdelta-\sum_{i=1}^k\frac{a_i\bxi_i^{\top}\bdelta}{a(a+a_i)}\bxi_i.
\end{eqnarray}
In other words, $\bbeta$ is in the space spanned by $\{\bdelta$, $\bxi_1$,..., $\bxi_k\}$. On the other hand, by (\ref{a2}), it is easy to see that the space spanned by eigenvectors of $\bSigma^{tot}$ corresponding to eigenvalues greater than $a$ is exactly the space spanned by  $\{\bdelta$, $\bxi_1$,..., $\bxi_k\}$. Therefore, $\bbeta$ is perpendicular to the $p-k-1$ dimensional eigenspace corresponding to eigenvalue $a$, i.e. $||\bU^{\top}\bbeta||_0\leq k+1$.
\hfill$\blacksquare$

Before proving Theorem 2, we need a couple of results on the eigenvalues and eigenspaces of hermitian/symmetric matrices.

\begin{lemma}\citep{weyl1912}\label{lemma2}
If $A$ and $B$ are symmetric $p\times p$ matrices that differ by a matrix of rank at most $r$, then their eigenvalues (in descending order) $\{\alpha_j\}_{1\leq j\leq p}$ and $\{\gamma_j\}_{1\leq j\leq p}$ satisfy \[\alpha_{j+r}\leq\gamma_{j} \quad \text{and}\quad \gamma_{j+r}\leq\alpha_{j} \quad \text{for}\quad 1\leq j,\text{ }j+r\leq p.\]
In particular, if $r=1$ and $A\geq B$, it implies an interlacing property
\[\alpha_1\geq\gamma_1\geq\alpha_2\geq\cdots\geq\alpha_p\geq\gamma_p.\]
\end{lemma}

\begin{lemma}\citep{davis1970rotation}\label{lemma3}
Let $A$ and $B$ be symmetric matrices with $A-B=H$ and eigenvalues $\{\alpha_j\}_{1\leq j\leq p}$ and $\{\gamma_j\}_{1\leq j\leq p}$, respectively. If there exist a subset $\mathcal{S}\subset\{1,...,p\}$, an interval $[s,t]$ and a positive constant $z$, such that  $\alpha_j,\gamma_j\in[s,t]$ when $j\in \mathcal{S}$ and  $\alpha_j,\gamma_j\in(-\infty, s-z]\cup[t+z,\infty)$ when $j\notin \mathcal{S}$, then $||P-Q||\leq ||H||/z$, where $P$ and $Q$ are projection matrices to the subspaces spanned by eigenvectors corresponding to $\{\alpha_j\}_{ j\in\mathcal{S}}$ and $\{\gamma_j\}_{j\in\mathcal{S}}$, respectively.
\end{lemma}

The following lemmas are crucial in the proof of Theorem 2.
\begin{lemma}\label{lemma4}
Under Condition 2, if $\bdelta\in\bW_1$, then the eigenvalues of $\bSigma^{tot}$ satisfy
\begin{eqnarray}\label{a4}
\eta_1\geq\eta_2\geq\cdots\geq\eta_k\geq \eta_{k+1}+d>\eta_{k+1}\geq\cdots\geq\eta_p;
\end{eqnarray}
otherwise,
\begin{eqnarray}\label{a5}
\eta_1\geq\eta_2\geq\cdots\geq \eta_{k+1}\geq \eta_{k+2}+d\frac{\rho||\bdelta_2||^2_2}{d+\rho||\bdelta||^2_2}-\epsilon \geq \eta_{k+2}\geq\cdots\geq\eta_p.
\end{eqnarray}
\end{lemma}
{\bf Proof of Lemma \ref{lemma4}}: Recall that $\{\lambda_j\}_{1\leq j\leq p}$ are eigenvalues of $\bSigma$ in descending order and $\bxi_j$ is the eigenvector corresponding to $\lambda_j$. $\bW_1$ and $\bW_2$ are linear spaces spanned by $\{\bxi_j\}_{1\leq j\leq k}$ and $\{\bxi_j\}_{k+1\leq j\leq p}$, respectively. $\bdelta=\bdelta_1+\bdelta_2$ with $\bdelta_m\in\bW_m$, $m=1,2$.

If $\bdelta\in \bW_1$, then $\bdelta\perp\bxi_j$ for $k+1\leq j\leq p$. Therefore, $\{\bxi_j\}_{k+1\leq j\leq p}$ are eigenvectors of $\bSigma^{tot}=\bSigma+\rho\bdelta\bdelta^{\top}$ as well, and the corresponding eigenvalues satisfy $\eta_j=\lambda_j$ for $k+1\leq j\leq p$. Moreover, by Lemma \ref{lemma2}, $\eta_1\geq\lambda_1\geq\eta_2\geq\cdots\geq\eta_k\geq\lambda_k$. Thus, Condition 2 implies
\begin{eqnarray*}
\eta_1\geq\eta_2\geq\cdots\geq\eta_k\geq \eta_{k+1}+d>\eta_{k+1}\geq\cdots\geq\eta_p.
\end{eqnarray*}

If $\bdelta\notin\bW_1$, i.e., $\bdelta_2\neq0$, define $\bW=\bW_1\oplus\bdelta_2$. For all $\bw\in\bW$, with $||\bw||_2=1$, we may write $\bw=\bw_1+\bw_2$ where $\bw_1\in\bW_1$ and $\bw_2=c\bdelta_2\in\bW_2$.  It follows that
\begin{eqnarray*}
\bw^{\top}\bSigma^{tot}\bw&=&\bw^{\top}\bSigma\bw+\rho\bw^{\top}\bdelta\bdelta^{\top}\bw\\
&=&\bw_1^{\top}\bSigma\bw_1+\bw_2^{\top}\bSigma\bw_2+\rho\left((\bw_1^{\top}+\bw_2^{\top})(\bdelta_1+\bdelta_2)\right)^2\\
&=&\bw_1^{\top}\bSigma\bw_1+\bw_2^{\top}\bSigma\bw_2+\rho\left(\bw_1^{\top}\bdelta_1+\bw_2^{\top}\bdelta_2\right)^2\\
&\geq&\lambda_k||\bw_1||^2_2+\lambda_p||\bw_2||^2_2+\rho\left(\bw_1^{\top}\bdelta_1+\bw_2^{\top}\bdelta_2\right)^2\\
&\geq&\lambda_p+d||\bw_1||^2_2+\rho\left(|\bw_1^{\top}\bdelta_1|-|\bw_2^{\top}\bdelta_2|\right)^2\\
\end{eqnarray*}

It is easy to see that
\[\inf\left\{\rho\left(|\bw_1^{\top}\bdelta_1|-|\bw_2^{\top}\bdelta_2|\right)^2\right\}=\left\{
                                                                               \begin{array}{ll}
                                                                                 0, & \hbox{if } ||\bw_1||_2\geq ||\bdelta_2||_2/||\bdelta||_2; \\
                                                                                 \rho(||\bw_2||_2||\bdelta_2||_2-||\bw_1||_2||\bdelta_1||_2)^2, & \hbox{if } ||\bw_1||_2< ||\bdelta_2||_2/||\bdelta||_2.
                                                                               \end{array}
                                                                             \right.\]

Therefore, if $||\bw_1||_2\geq ||\bdelta_2||_2/||\bdelta||_2$,
\[\bw^{\top}\bSigma^{tot}\bw\geq\lambda_p+d||\bw_1||^2_2\geq\lambda_p+d\frac{||\bdelta_2||^2_2}{||\bdelta||^2_2};\]
if $||\bw_1||_2< ||\bdelta_2||_2/||\bdelta||_2$,
\begin{eqnarray*}
\bw^{\top}\bSigma^{tot}\bw&\geq& \lambda_p+d||\bw_1||^2_2+ \rho(||\bw_2||_2||\bdelta_2||_2-||\bw_1||_2||\bdelta_1||_2)^2\\
&\geq&\lambda_p+d||\bw_1||^2_2+\rho(||\bdelta_2||_2-||\bw_1||_2 ||\bdelta||_2)^2\\
&\geq&\lambda_p+d\frac{\rho||\bdelta_2||^2_2}{d+\rho||\bdelta||^2_2}.
\end{eqnarray*}

Overall, we have
\begin{eqnarray}\label{a6}
\bw^{\top}\bSigma^{tot}\bw\geq\lambda_p+\tilde d\quad\text{for all}\quad \bw\in\bW,
\end{eqnarray}
where $\tilde d=d\frac{\rho||\bdelta_2||^2_2}{d+\rho||\bdelta||^2_2}$. Since $\dim\bW=k+1$, (\ref{a6}) implies that there are $k+1$ eigenvalues that are greater than $\lambda_p+\tilde d$ for $\bSigma^{tot}$. Together with Lemma \ref{lemma2}, we conclude
\begin{eqnarray*}
\eta_1\geq\eta_2\geq\cdots\geq\eta_k\geq \eta_{k+1}\geq \lambda_p+\tilde d>\lambda_{k+1}\geq \eta_{k+2}\geq\cdots\geq\eta_p.
\end{eqnarray*}
which leads to (\ref{a5}).\hfill$\blacksquare$

Similarly, we have
\begin{lemma}\label{lemma5}
Under Condition 1, if $\bdelta\in\bW_1$, then the eigenvalues of $\bSigma^{tot}$ satisfy
\begin{eqnarray}\label{a7}
\eta_1\geq\eta_2\geq\cdots\geq\eta_k\geq \eta_{k+1}+d>\eta_{k+1}=\cdots=\eta_p;
\end{eqnarray}
otherwise,
\begin{eqnarray}\label{a8}
\eta_1\geq\eta_2\geq\cdots\geq\eta_k\geq \eta_{k+1}\geq \eta_{k+2}+d\frac{\rho||\bdelta_2||^2_2}{d+\rho||\bdelta||^2_2} \geq \eta_{k+2}=\cdots=\eta_p.
\end{eqnarray}
\end{lemma}

{\bf Proof of Lemma \ref{lemma5}}: The only difference is that the last $p-k-1$ eigenvalues are equal, which is implies by Lemma \ref{lemma2} and the fact that $\lambda_{k+1}=\lambda_{k+2}=\cdots=\lambda_p$. \hfill$\blacksquare$

{\bf Proof of Theorem 2}: Again, let $\bxi_j$ be the eigenvector of $\bSigma$ corresponding to $\lambda_j$ for $1\leq j \leq p$. $a=\lambda_p$ and $a_j=\lambda_j-\lambda_p$.

{\bf Part I:} $\bdelta\in\bW_1$ implies $\bdelta\perp\bxi_j$ for $k<j\leq p$, so the eigenvectors $\{\bxi_j\}_{k<j\leq p}$ are also eigenvectors of $\bSigma^{tot}$. Write $\bU=\left(\bU_1\text{ }\bU_2\right)$ where $\bU_2$ is submatrix of $\bU$, consisting of right $p-k$ columns. Then $\bU_2 =\left(\bxi_{k+1},\cdots,\bxi_p\right)$. Therefore,
\[\bU_2^{\top}\bbeta=\bU^{\top}_2\bSigma^{-1}\bdelta=\bD^{-1}_2\bU_2^{\top}\bdelta=\bzero, \]
where $\bD_2=diag(\lambda_{k+1},...,\lambda_p)$.

{\bf Part II:} Under Condition 2, we can write $\bSigma=\bSigma_0+\bDelta$ where $\bSigma_0=a\bI+\sum_{j=1}^ka_j\bxi_j\bxi_j^{\top}$ and $\bDelta=\sum_{j=k+1}^pa_j\bxi_j\bxi_j^{\top}$. Thus, $\bSigma_0$ satisfies Condition 1, and $\bDelta$ is a semipositive matrix with maximal eigenvalue less than $\epsilon$. Define
\[\bSigma^{tot}_0=\bSigma_0+\rho\bdelta\bdelta^{\top} \quad\text{and}\quad \bSigma^{tot}=\bSigma+\rho\bdelta\bdelta^{\top}.\]
And let $\{\eta_{0j}\}_{1\leq j\leq p}$ and $\{\eta_{j}\}_{1\leq j\leq p}$ be their eigenvalues, in the descending order, respectively. Moreover, let $\bV$ and $\bU$ be orthogonal matrices such that
\[\bV^{\top}\bSigma^{tot}_0\bV=\bD_0 \quad\text{and}\quad \bU^{\top}\bSigma^{tot}\bU=\bD,\]
where $\bD_0=diag(\eta_{01},...,\eta_{0p})$ and $\bD=diag(\eta_1,...,\eta_p)$.

Here is the strategy of the proof. By Theorem 1, $\bV^{\top}\bSigma_0^{-1}\bdelta$ is sparse so its $\ell_1$-norm can be well controlled. Because of the results on the separated eigenvalues (Lemmas \ref{lemma4} and \ref{lemma5}),  we can show $\bU^{\top}\bbeta$ is similar to $\bV^{\top}\bSigma_0^{-1}\bdelta$ using Lemma \ref{lemma3}. Therefore, the $\ell_1$-norm can be controlled as well.

Write $\bU=\left(\bU_1\text{ }\bU_2\right)$ and $\bV=\left(\bV_1\text{ }\bV_2\right)$ where $\bU_2$ and $\bV_2$ are submatrices of $\bU$ and $\bV$ respectively, consisting of right $p-k-1$ columns. Note that
\[||\bU^{\top}\bbeta||_1=||\bU^{\top}_1\bbeta||_1+||\bU^{\top}_2\bbeta||_1,\]
where $||\bU^{\top}_1\bbeta||_1 \leq \sqrt{||\bU^{\top}_1\bbeta||_0\cdot||\bU^{\top}_1\bbeta||^2_2}\leq\sqrt{k+1}||\bbeta||_2$. So it is crucial to control $||\bU^{\top}_2\bbeta||_1$. From the proof of Theorem 1, we see that $\bV_2^{\top}\bSigma_0^{-1}\bdelta=\bzero$. Hence
\begin{eqnarray*}
||\bU_2^{\top}\bbeta||_2&=&||\bU_2^{\top}\bSigma^{-1}\bdelta-\bU_2^{\top}\bSigma_0^{-1}\bdelta+\bU_2^{\top}\bSigma_0^{-1}\bdelta||_2\\
&\leq&||\bU_2^{\top}\bSigma^{-1}\bdelta-\bU_2^{\top}\bSigma_0^{-1}\bdelta||_2+||\bU_2^{\top}\bSigma_0^{-1}\bdelta||_2\\
&\leq&||\bU_2^{\top}\bSigma^{-1}\bdelta-\bU_2^{\top}\bSigma_0^{-1}\bdelta||_2+\sqrt{||\bU_2^{\top}\bSigma_0^{-1}\bdelta||_2^2-||\bV_2^{\top}\bSigma_0^{-1}\bdelta||_2^2}\\
&\leq&||\bSigma^{-1}-\bSigma_0^{-1}||\cdot||\bdelta_2||_2+\sqrt{\bdelta^{\top}(\bSigma_0^{-1})^{\top}\left(\bU_2\bU_2^{\top}-\bV_2\bV_2^{\top}\right)\bSigma_0^{-1}\bdelta} \\
&=&S_1+S_2
\end{eqnarray*}
and
\[||\bSigma^{-1}-\bSigma_0^{-1}||=\lambda_p^{-1}-\lambda_{k+1}^{-1}=a^{-1}-(a+a_{k+1})^{-1}=\frac{a_{k+1}}{a(a+a_{k+1})}\leq\frac{\epsilon}{a^2}.\]
Thus, $S_1\leq \frac{\epsilon}{a^2}||\bdelta_2||_2\leq \frac{\epsilon(a+\epsilon)}{a^2}||\bbeta||_2$.

To control $S_2$, we have to show that the spaces spanned by column vectors of $\bV_2$ and $\bU_2$ are close to each other.
By Lemmas \ref{lemma4} and \ref{lemma5}, we have
\[\eta_1\geq\eta_2\geq\cdots\geq\eta_k\geq \eta_{k+1}\geq \eta_{k+2}+\tilde d-\epsilon \geq \eta_{k+2}\geq\cdots\geq\eta_p,\]
\[\eta_{01}\geq\eta_{02}\geq\cdots\geq\eta_{0k}\geq \eta_{0,k+1}\geq \eta_{0,k+2}+ \tilde d \geq \eta_{0,k+2}=\cdots=\eta_{0p},\]
where $\tilde d=d\frac{\rho||\bdelta_2||^2_2}{d+\rho||\bdelta||^2_2}$. Moreover, by Lemma \ref{lemma2}, $\eta_{k+2}\leq\lambda_{k+1}\leq\lambda_p+\epsilon=a+\epsilon$,  $\eta_{0,k+2}=\lambda_p=a$. On the other hand, $\eta_{k+1}\geq  \eta_{k+2}+\tilde d-\epsilon\geq a+\tilde d-\epsilon$, $\eta_{0,k+1}\geq \eta_{0,k+2}+ \tilde d=a+\tilde d$.

By Lemma \ref{lemma3}, $||\bU_2\bU_2^{\top}-\bV_2\bV_2^{\top}||\leq ||\bDelta||/(\tilde d-2\epsilon)\leq\epsilon/(\tilde d-2\epsilon)$.

$||\bSigma_0^{-1}\bdelta||_2=||\bSigma_0^{-1}\bSigma\bbeta||_2\leq ||\bSigma_0^{-1}\bSigma||\cdot ||\bbeta||_2\leq\frac{a+\epsilon}{a}||\bbeta||_2$. Thus, $S_2\leq \sqrt{\frac{\epsilon}{\tilde d-2\epsilon}}\frac{a+\epsilon}{a}||\bbeta||_2$. Therefore,
$||\bU_2^{\top}\bbeta||_2\leq\frac{a+\epsilon}{a}(\frac{\epsilon}{a} +\sqrt{\frac{\epsilon}{\tilde d-2\epsilon}})||\bbeta||_2$. $||\bU_2^{\top}\bbeta||_1\leq\sqrt{p-k-1}\frac{a+\epsilon}{a}(\frac{\epsilon}{a} +\sqrt{\frac{\epsilon}{\tilde d-2\epsilon}})||\bbeta||_2$.

Finally, $||\bU^{\top}\bbeta||_1/||\bbeta||_2\leq\sqrt{k+1}+\sqrt{p-k-1}\frac{a+\epsilon}{a}(\frac{\epsilon}{a} +\sqrt{\frac{\epsilon}{\tilde d-2\epsilon}})$. \hfill$\blacksquare$

{\bf Proof of Theorem~\ref{thm3}}.  Let $\bV_1$ be a matrix whose columns vectors are the eigenvectors corresponding to the nonvanishing eigenvalues of the matrix $\bA = \sum_{i=1}^k \lambda_i \bxi_i \bxi_i^{\top} + \rho \bdelta \bdelta^{\top}$.   Recall $\lambda_i(\bB)$ be the $i^{th}$ largest eigenvalue of a symmetric matrix $\bB$.
Then, by Lemma~\ref{lemma3},
$$
    \| \bU_1 - \bV_1 \| \leq \frac{\| \bSigma^{tot} - \bA \|}{\lambda_{k+1}(\bA) - \lambda_{k+2}(\bSigma^{tot})} =
    \frac{\lambda_{k+1}}{\lambda_{k+1}(\bA) - \lambda_{k+2}(\bSigma^{tot})}.
$$
By Lemma~\ref{lemma2}, $\lambda_{k+2}(\bSigma^{tot}) \leq \lambda_{k+1}$.  Hence,
$$
\| \bU_1 - \bV_1 \| \leq \frac{\lambda_{k+1}}{\lambda_{k+1}(\bA) - \lambda_{k+1}} \leq \frac{1}{a-1},
$$
Let $\bV_2$ be the eigenvectors that are orthogonal to $\bV_1$.  Then, $\bV_2^{\top} \bdelta = 0$, since the columns of $\bV_1$ are the linear combinations of $\bdelta$ and $\{\bxi_i\}_{i=1}^k$.  Consequently, $\|\bV_1^{\top} \bdelta \|_2 = \|\bdelta\|_2$ and
$$
\|\bU_1^{\top} \bdelta \|_2 = \| \bV_1^{\top} \bdelta + (\bU_1 - \bV_1)^{\top} \bdelta \|_2 \geq \|\bdelta\|_2 - \|\bU_1 - \bV_1\| \| \bdelta \|_2 = \frac{a-2}{a-1} \|\bdelta \|_2.
$$
The second conclusion follows directly from the fact that
$\|\bU_1^{\top} \bSigma \bU_1\| \leq \|\bSigma\| = \lambda_1$.   \hfill$\blacksquare$

\bibliographystyle{biometrika}
\bibliography{ROSEreference}
\end{document}